\documentclass[11pt,a4paper,review]{article}

\usepackage[T1]{fontenc}
\usepackage[utf8]{inputenc}
\usepackage{amsmath,amssymb,amsthm}
\usepackage{mathtools}
\usepackage{booktabs}
\usepackage{array}
\usepackage{geometry}
\usepackage{xcolor}
\usepackage{setspace}
\usepackage[round,authoryear]{natbib}
\usepackage[hidelinks]{hyperref}
\usepackage{enumitem}
\usepackage{graphicx}
\usepackage{tikz}
\usepackage{needspace}
\usepackage{algorithm}
\usepackage{algpseudocode}
\usepackage{longtable}

\theoremstyle{plain}
\newtheorem{theorem}{Theorem}
\newtheorem{proposition}{Proposition}
\newtheorem{lemma}{Lemma}
\newtheorem{corollary}{Corollary}

\theoremstyle{definition}
\newtheorem{definition}{Definition}
\newtheorem{assumption}{Assumption}

\newtheorem{remark}{Remark}

\newcommand{\N}{\mathcal{N}}
\newcommand{\G}{\mathcal{G}}

\newcommand{\gK}{g^{\mathrm{c}}}
\newcommand{\gKm}{g^{\mathrm{c}-}}
\newcommand{\gE}{g^{\emptyset}}
\newcommand{\gS}{g^{*}}
\newcommand{\gC}{g^\circ}
\newcommand{\gMP}[1]{g_{#1}}

\newcommand{\gA}[1]{g^{(#1)}}

\newcommand{\dist}[3]{d_{#1}(#2,#3)}
\newcommand{\ind}[1]{\mathbf{1}\{#1\}}

\newcommand{\Ben}{B}
\newcommand{\W}{W}

\newcommand{\Eff}{\mathcal{E}}
\newcommand{\PS}{\mathcal{P}}

\newcommand{\rev}[1]{#1}
\newcommand{\conf}[1]{#1}
\newenvironment{confblock}{}{\ignorespacesafterend}
\newenvironment{revblock}{}{\ignorespacesafterend}
\newenvironment{redblock}{}{\ignorespacesafterend}
\newcommand{\redtxt}[1]{#1}

\DeclareMathOperator*{\argmax}{arg\,max}

\title{\bfseries Optimal Network Design under Node Failures%
}

\author{%
  Michele Lombardi\thanks{\emph{Corresponding author}. University of Liverpool Management School, University
  of Liverpool, UK, and Department of Economics and Statistics, University of
  Naples Federico II, Italy. Email:
  \texttt{michele.lombardi@liverpool.ac.uk}.}
\and
  Francesco Ciardiello\thanks{Department of Economics and Statistics,
  University of Salerno, Italy. Email: \texttt{fciardiello@unisa.it}.}
  \and
  Enjun Liu\thanks{Department of Economics, International Business School
  Suzhou, Xi'an Jiaotong-Liverpool University, Suzhou, China. Email:
  \texttt{Enjun.Liu23@student.xjtlu.edu.cn}.}
}

\date{\today}

\begin{document}
\maketitle

\onehalfspacing

\begin{abstract}
\noindent\rev{Networks are frequently constructed before it is determined which nodes will persist. We address the design problem in scenarios where nodes fail independently, links are costly and must be established in advance, and the benefits to surviving nodes diminish with distance. Examples include supplier, interbank, and infrastructure networks. Among all possible networks, the complete, star, and empty configurations are each welfare-optimal within specific ranges of link costs. However, with four or more nodes, failures introduce a range of costs in which every optimal network contains more links than the star but fewer than the complete network. At the lower end of this range, we show that the optimal networks are exactly those obtained by removing a matching from the complete network. We develop an exact algorithm for complete multipartite networks. It finds an optimal network by dynamic programming in a number of operations that grows with the fourth power of the number of nodes, although the class itself grows superpolynomially. For networks with up to six nodes, restricting the design to this class results in no loss under explicit conditions. We also characterize pairwise-stable networks within this class and establish their existence for every level of link cost. 
Finally, we quantify the cost of misjudging failure risk. Overestimating node survival leads to overbuilding and can yield lower welfare than building nothing, whereas underestimating it never does. Under uncertain failure risk, design should therefore favor conservative reliability estimates.}


\end{abstract}

\section{Introduction}
Networks are often designed before their future operating conditions are known. Links must be established, installed, or contracted in advance, while the agents, facilities, or organizations that make those connections valuable may subsequently become unavailable. This raises \rev{a central design question of which network} should be formed when connectivity generates benefits, links are costly, and nodes may \rev{fail.}

Reliability has long played a central role in network design, particularly in communication, transportation, infrastructure, and distribution networks. Traditional reliability problems typically focus either on assessing whether a network can remain connected or continue providing service after component failures or on designing networks that satisfy prescribed reliability requirements.

Our approach takes a different, more economic perspective. \rev{Rather than imposing a reliability constraint or permitting post-failure repairs, we assume that links are chosen and paid for before node survival is realized. Nodes then survive independently, and connectivity benefits arise from the residual network. Reliability thus enters the design problem endogenously, through the expected benefits of connectivity rather than an externally imposed target.} 

\rev{Link costs are sunk, so links incident to failed nodes are paid for but yield no connectivity benefits. Node failures may also} eliminate indirect connections or increase the distances among surviving nodes. Consequently, network design must balance certain \rev{link} costs against uncertain connectivity benefits. Dense networks require greater ex ante investment but provide alternative paths, whereas sparse networks incur lower costs but depend more on intermediaries. The star topology exemplifies this trade-off\rev{: it} connects all nodes with relatively few links; however, if the central node fails, the remaining peripheral nodes become disconnected from each other.

\rev{This trade-off arises in settings where relationships or infrastructure must be established before future availability is known. For example, firms establish supplier relationships before knowing which suppliers may continue operating after disruptions. Organizations invest in collaborative arrangements before knowing whether participants will remain committed. Similarly, financial institutions form costly bilateral relationships while counterparties’ future availability remains uncertain. Infrastructure links are installed before it is clear which facilities may become unavailable.}

\rev{We capture this timing by separating network formation from network realization. Links are formed and paid for first, and only afterward is it revealed which nodes survive. Formally, we build on the connections model of \citet{JW1996}. Agents benefit from direct and indirect connections, with benefits decreasing as the distance between agents increases. Forming direct links is costly. Each node survives independently with probability $p$, and benefits depend on the surviving network.  When $p=1$, the model reduces to the standard connections model. For $p<1$, however, network structure determines not only distances but also the resilience of indirect connectivity to node failures.}

\rev{In this model, we ask the following questions. First, for which link costs do the complete, star, and empty networks remain optimal, and which networks are optimal in between? Second, can an optimal network be computed efficiently when the number of nodes is large, and how much is lost by restricting the search to a tractable class of networks? Third, what does it cost a planner to misjudge the risk of failure? Finally, which networks are pairwise stable when agents form links themselves, does such a network always exist, and can it be optimal?}

\paragraph{Contribution of the \rev{paper}.}\mbox{} The paper makes three contributions: an exact method for welfare-optimal design within complete multipartite networks, a broader characterization of welfare-optimal networks, and an analysis of pairwise-stable outcomes under decentralized formation.

\begin{redblock}
\textbf{Exact design.}
Our main contribution is an exact, polynomial-time method for
welfare-optimal network design within the class of complete
multipartite networks. This class contains the empty network,
the star, the complete network, and a broad family of intermediate
symmetric and asymmetric networks, each represented by a partition
of the nodes. We derive expected benefits in closed form, characterize the benefit frontier, which records maximal benefits at each feasible link count,
and use dynamic programming to solve the restricted welfare-maximization problem in $O(n^4)$ arithmetic operations. The number of complete multipartite networks grows superpolynomially, yet both the maximum welfare \rev{attainable within this class} and a \rev{network attaining it} are found in polynomial time.

\rev{We put the algorithm to three uses. It traces how the optimal network
changes with link costs and survival probabilities, from asymmetric to
increasingly balanced multipartite structures. It quantifies the cost of
misjudging failure risk.} Overestimating node survival leads to excessive
link formation and can result in negative welfare, \conf{whereas underestimating it never does.} \conf{It also provides a benchmark for simple heuristics. In the instances tested, greedy link addition finds the optimum only when the complete network is optimal. Where it is not, greedy link deletion misses the optimum in more than a third of the instances, and a single local-search run misses it in about half.}

\textbf{Structure of optimal networks.}
Over the unrestricted class of networks on a fixed set of $n$ nodes,
node failures substantially enrich the classical characterization in
terms of complete, star, and empty networks, since intermediate
networks, with link counts between those of the star and the complete
network, can also be welfare-optimal. At the same time, the complete,
star, and empty networks remain welfare-optimal over explicit ranges
of link costs. At a critical switching cost, the complete network is co-optimal with every network obtained by removing a matching, namely a set of pairwise nonincident links. Immediately above this threshold, only the
network obtained by removing a maximum-cardinality matching remains
optimal, so the multiplicity of optimal networks gives way to a more
selective structure as link costs increase. Restricting attention to complete multipartite networks may exclude some welfare-optimal networks from the unrestricted problem, but for networks with up to six nodes the loss is limited. The restricted and unrestricted sets of welfare-optimal networks coincide over a substantial portion of the parameter space, and we give sufficient conditions guaranteeing this coincidence.

\textbf{Decentralized formation.}
\rev{Having characterized socially optimal networks, we next ask whether
decentralized link formation can sustain them. We therefore let the nodes
act as economic agents.} Each agent may unilaterally sever an existing link,
while two agents may jointly form a new one, and we adopt pairwise
stability as the equilibrium concept. We characterize precisely which complete multipartite networks are pairwise stable against such deviations, even when the deviations lead to networks outside the class, and we prove that a pairwise-stable network exists for every admissible parameter configuration. Stability and welfare optimality can nevertheless be incompatible. Individual decisions account for the benefits of the agents directly involved, whereas the social value of a link also includes its
contribution to indirect connectivity among other agents. Hence, the absence of profitable individual deviations does not by itself guarantee welfare optimality.
\end{redblock}

\begin{revblock}
The remainder of the paper is organized as follows. Section \ref{sec:related} reviews the related literature, and Section \ref{sec:model} presents the model. Section \ref{sec:classes} introduces complete multipartite and benchmark networks and formulates the restricted and unrestricted optimization problems and the benefit frontier. Section \ref{sec:efficiency} characterizes welfare-optimal networks in the unrestricted class. Section \ref{sec:computation} studies the restricted problem, compares its solutions with unrestricted optima for small networks, presents a dynamic programming algorithm (Section \ref{sec:algorithm}), and reports numerical results. Section \ref{sec:stability} analyzes pairwise stability, and Section \ref{sec:extensions} concludes. All proofs are collected in Appendix \ref{app:proofs}.
\end{revblock}

\section{\texorpdfstring{\rev{Related Literature}}{Related Literature}}\label{sec:related}
Our analysis lies at the intersection of three strands of literature\rev{, namely} network reliability and structural reliability
synthesis, reliable and survivable network design, and the economics of network formation. \rev{These strands share} an interest in \rev{network design} under uncertainty, but differ in their objectives, failure mechanisms, and formation decisions.

\begin{revblock}
The failure mechanism in our model is that of residual node connectedness reliability, in which edges are perfectly reliable, nodes fail independently, and the surviving nodes induce the residual network \citep{BSS1991}. That measure is the probability that the residual network is connected. Computing it is NP-hard in general \citep{SSS1991}, although efficient algorithms exist for particular graph classes \citep{CSSS1993}. Our objective differs from it in that we value each surviving pair according to its distance in the residual network.
Node failures may therefore reduce welfare even without disconnecting the network, by lengthening the distance between surviving nodes.
\end{revblock}

\begin{revblock}
A related literature on reliability synthesis asks which graph, among all graphs with a given number of nodes and edges, maximizes residual node connectedness reliability for every survival probability. Such uniformly best graphs need not
exist, but several families of complete multipartite graphs with nearly equal
cell sizes have this property \citep{GJL1994,Liu2000,YSM2010}. Those results fix
the number of edges and concern connectedness alone. In our model the number of links is chosen, links are costly, and benefits depend on distance, so the optimal cell sizes vary with the link cost and need not be balanced.
\end{revblock}

\begin{revblock}
The Operations Research literature on reliable and survivable network design also studies the choice of a network under uncertainty. One line of work minimizes cost subject to a reliability requirement, such as a lower bound on all-terminal reliability \citep{AltiparmakDengiz2009}, or allocates resources to reduce failure probabilities within a design problem \citep{DesaiSen2010}. Survivable network design typically imposes connectivity or service requirements
that must hold after disruptions \citep{Balakrishnan2013,GouveiaLeitner2017},
and stochastic variants may allow recourse once uncertainty is realized
\citep{LjubicMutzelZey2017}. Our model has neither a reliability constraint nor
post-failure adjustment. All links are formed and paid for ex ante, and welfare
depends on the resulting residual network. Models of facility and hub location
under disruptions \citep{Cui2010,KimOKelly2009,Li2022} address a related
question, but their decisions concern mainly where to locate facilities and how
to assign or back up demand.
\end{revblock}

The economic foundation of our analysis is the
connections model of \citet{JW1996}, recovered when
$p=1$. Introducing node \rev{failures} gives links an
additional role\rev{. Besides} creating or shortening
connections, they may preserve connectivity when
intermediaries disappear.

\begin{redblock}
Related work on network formation examines reliability under different failure mechanisms and formation rules. \citet{BG2000b} study unreliable links, \citet{Billand2016} consider exogenous node failures, and other contributions analyze strategic attacks on networks \citep{DG2013,GoyalVigier2014}. Our model brings together independent node failures and bilateral link formation, allowing us to compare welfare-optimal networks with outcomes sustained by decentralized formation. Finally, our work is inspired by \citet{LJG2025}, who study the same environment in a financial-network setting. They obtain initial results on the efficiency and pairwise stability of three benchmark networks: the complete, star, and empty networks. In this paper, however, we focus more systematically on network design and computation.
\end{redblock}

\section{The Model}\label{sec:model}

Let $\N=\{1,2,\dots,n\}$ be a finite set of nodes or agents, with $n\ge 3$. Throughout the paper, we use the terms \textit{nodes} and \textit{agents} interchangeably, referring to nodes when describing graph structure from a normative point of view, and to agents when discussing individual benefits and strategic decisions.
A \emph{link} is an unordered pair $ij=\{i,j\}$ with $i,j\in\N$ and $i\ne j$, and represents a costly bilateral relationship that both parties must agree to maintain. Write
$K_\N=\bigl\{ij:i,j\in\N,\;i\ne j\bigr\}$
for the set of all $\binom{n}{2}$ possible links.

\begin{definition}[Graph]
A graph is a subset $g\subseteq K_\N$.
\end{definition}

We write $ij\in g$ when the link $ij$ is present in $g$, and $|g|$ for the number of links. For $ij\notin g$ let $g+ij=g\cup\{ij\}$, and for $ij\in g$ let $g-ij=g\setminus\{ij\}$. The neighborhood of $i$ is $N_i(g)=\{j\in\N:ij\in g\}$, its degree is $\deg_i(g)=|N_i(g)|$, and the \emph{degree sequence} is the vector of degrees arranged in non-increasing order. A path between $i$ and $j$ in $g$ is a sequence $i=v_0,v_1,\dots,v_m=j$ of distinct nodes such that $v_{\ell-1}v_\ell\in g$ for every $\ell$\rev{. Its} length is $m$. The graph $g$ is connected if every pair of distinct nodes is joined by a path.

\begin{assumption}[Failure]\label{ass:failure}
Each node survives with probability $p\in(0,1)$, independently across nodes. Failure is independent of the graph and of the actions of any agent.
\end{assumption}
The boundary case $p=1$ provides a natural benchmark. In the absence of node failures, the model reduces to the standard connections model of \citet{JW1996}. We focus on $p\in(0,1)$, since our interest lies in the case of uncertain node survival and in the resulting implications for graph design, rather than in revisiting the deterministic benchmark.
\rev{Throughout, we write $q=1-p$ for the probability that a node fails.}

Let $S\subseteq\N$ denote the realized set of surviving nodes. Under Assumption \ref{ass:failure}, its probability is
\begin{equation}\label{eq:lambda}
\lambda(S)=p^{|S|}\rev{q^{\,n-|S|}},
\qquad
\sum_{S\subseteq\N}\lambda(S)=1.
\end{equation}
A failed node exits the graph together with all of its links, so the graph operating in state $S$ is the induced subgraph
$g|_S=\{ij\in g:i\in S,\;j\in S\}.$
Write $\dist{g|_S}{i}{j}$ for the length of a shortest path between $i$ and $j$ in $g|_S$, with $\dist{g|_S}{i}{j}=\infty$ when no such path exists. Distances are therefore computed after the failure shock\rev{, so} the disappearance of an intermediate node can increase the distance between surviving nodes or disconnect them.

\begin{assumption}[Benefits and costs]\label{ass:benefit}
There is a decay factor $\delta\in(0,1)$ and a \rev{link} cost $c>0$. A node $i$ that survives collects $\delta^d$ from every surviving node at distance $d<\infty$ from $i$ in $g|_S$, with the convention $\delta^\infty=0$. The \rev{link} cost $c$ is borne by each endpoint of every link, is sunk at the formation stage, and does not depend on which nodes survive.
\end{assumption}

The restriction $\delta\in(0,1)$ ensures that indirect connections generate positive but strictly decreasing benefits with distance. The boundary cases $\delta=0$ and $\delta=1$ would eliminate, respectively, the value of all connections or the role of distance in determining connection benefits.

\begin{definition}[Expected payoff]\label{def:payoff}
The expected payoff of node $i$ in graph $g$ is
\begin{equation}\label{eq:payoff}
\Pi_i(g)
=
\underbrace{
\sum_{\substack{S\subseteq\N\\ i\in S}}
\lambda(S)
\sum_{j\in S\setminus\{i\}}
\delta^{\,\dist{g|_S}{i}{j}}
}_{\textstyle B_i(g)}
-
\deg_i(g)c,
\end{equation}
where $B_i(g)$ is the expected connection benefit of node $i$.
\end{definition}

The decay factor captures the value of indirect connections. Following \citet{JW1996}, a direct connection generates $\delta$, a two-step connection $\delta^2$, and, more generally, a connection at distance $d$ generates $\delta^d$. \rev{That is,} $\delta$ measures the fraction of value retained at each additional step of intermediation. As $\delta\to1$, distance becomes irrelevant and only reachability matters\rev{, whereas as $\delta\to0$} only direct connections generate positive benefits. Failure and distance interact because an indirect connection is valuable only in states in which the nodes along the corresponding path survive.

The timing implicit in \eqref{eq:payoff} is as follows\rev{. First,} the graph is formed and link costs are incurred\rev{. Second,} the failure shock is realized\rev{. Third,} surviving nodes receive the benefits generated by the residual graph. Hence, link costs are sunk before \rev{failures are} realized, whereas connection benefits depend on $g|_S$. A failed node therefore pays its link costs but receives no connection benefit.

Since the link cost is borne by both endpoints, each link generates an aggregate cost of $2c$, while its formation requires mutual consent.

\begin{definition}[Total benefit and welfare]\label{def:welfare}
The total expected connection benefit and the total expected welfare of a graph $g$ are
\begin{equation}\label{eq:welfare}
\Ben(g)=\sum_{i\in\N}B_i(g),
\qquad
\W(g)=\sum_{i\in\N}\Pi_i(g)
=\Ben(g)-2c|g|.
\end{equation}
\end{definition}

\section{\texorpdfstring{\rev{Network Classes and the Optimization Problems}}{Network Classes and the Optimization Problems}}\label{sec:classes}
\rev{This section introduces the classes of networks considered in the paper
and formulates the optimization problems studied in the remainder of the analysis.}

\subsection{\texorpdfstring{\rev{Networks and Complete Multipartite Networks}}{Networks and Complete Multipartite Networks}}\label{sec:networks}
\begin{definition}[Network]\label{def:network}
Two graphs $g,g'$ are isomorphic, written
$g\cong g'$, if there exists a permutation
$\pi$ of $\N$ such that $ij\in g$ if and only if
$\pi(i)\pi(j)\in g'$.
A \emph{network} is an equivalence class of graphs
under isomorphism. We denote by $\G(\N)$
the set of all networks on $\N$.
\end{definition}
Since expected aggregate benefits and welfare are
invariant under graph isomorphisms, they are
well defined on $\G(\N)$.
Throughout the paper, we identify a network with
any of its graph representatives whenever no
ambiguity arises. Operations involving individual
nodes or links are understood to be performed on
a specified representative.
We begin \rev{with an important class of networks,} the complete multipartite networks.
\begin{definition}[Complete multipartite network]\label{def:cmp}
A network $g\in\G(\N)$ is \emph{complete multipartite}
if $\N$ admits a partition into nonempty cells
$A_1,\ldots,A_r$ such that $ij\in g$ if and only if
$i$ and $j$ lie in different cells.

We denote by $\G^{\mathrm{mp}}(\N)\subseteq\G(\N)$
the set of complete multipartite networks.
For cell sizes $a_t=|A_t|$, let
$\gMP{a_1,\ldots,a_r}$ denote a representative
of the corresponding network,
so that $g\cong\gMP{a_1,\ldots,a_r}$.
Equivalently, $g$ is complete multipartite if and
only if any two distinct non-adjacent nodes have
identical neighborhoods.
\end{definition}
In the notation $\gMP{a_1,\ldots,a_r}$, the number
of indices specifies the number of cells, while
each index gives the size of the corresponding cell.
Their ordering is immaterial, as the cell sizes
uniquely determine the network.
For brevity, we write $a=(a_1,\ldots,a_r)\vdash n$
for an integer partition of $n$, and set
$\gMP{a}=\gMP{a_1,\ldots,a_r}$.
The class thus accommodates a range of intermediate
networks while allowing the welfare problem to be
expressed in terms of integer partitions of $n$.
\rev{The number of links, the expected benefits, and the payoffs of a complete
multipartite network have closed-form expressions.}
\begin{lemma}[Closed form]\label{lem:cmpform}
Let $g\cong\gMP{a_1,\dots,a_r}$ with
$\sum_t a_t=n$\rev{, and recall that $q=1-p$}. Then
\begin{equation}\label{eq:cmpL}
  |g|=\tfrac{1}{2}\Bigl(n^{2}-\sum_t a_t^{2}\Bigr)
      =\tfrac{1}{2}\sum_t a_t\,(n-a_t).
\end{equation}
\begin{equation}\label{eq:cmpB}
  \Ben(g)\;=\;2p^{2}\delta\,|g|
    \;+\;p^{2}\delta^{2}\sum_{t}a_t(a_t-1)
    \bigl(1-q^{\,n-a_t}\bigr).
\end{equation}
Moreover, for each node $i$ belonging to cell $t$,
the individual expected benefit and payoff are
\begin{equation}\label{eq:cmpayoff}
B_i(g)
=(n-a_t)p^{2}\delta
+(a_t-1)p^{2}\delta^{2}\bigl(1-q^{n-a_t}\bigr),
\qquad
\Pi_i(g)=B_i(g)-(n-a_t)c.
\end{equation}
\end{lemma}

\subsection{\texorpdfstring{\rev{Benchmark Networks}}{Benchmark Networks}}\label{sec:benchmarks}
For completeness, we introduce five benchmark networks that play a specific role throughout the \rev{paper.}

\begin{itemize}
\item The \emph{empty network} $\gE=\emptyset$,
with $|\gE|=0$, corresponding to the complete
multipartite network $\gMP{n}$.

\item The \emph{complete network} $\gK=K_\N$,
with $|\gK|=\binom{n}{2}$, corresponding to
$\gMP{1,\ldots,1}$.

\item The \emph{star network}
$\gS=\{hj:j\in\N\setminus\{h\}\}$,
centered at some node $h$, with $|\gS|=n-1$,
corresponding to $\gMP{1,n-1}$.

\item The \emph{complete network minus one link}
$\gKm$, obtained from $\gK$ by removing
a single link, with
$|\gKm|=\binom{n}{2}-1$,
corresponding to $\gMP{2,1,\ldots,1}$.

\item The \emph{cycle network} $\gC$,
in which all nodes belong to a single cycle,
with $|\gC|=n$.
\end{itemize}
The first four networks are complete multipartite,
whereas the cycle belongs to this class only for
$n=3$ or $n=4$. In particular, $\gC\cong\gMP{1,1,1}$
when $n=3$ and $\gC\cong\gMP{2,2}$ when $n=4$.
There are $n$ labeled stars, which are isomorphic\rev{, and}
we refer to them collectively as the star network.
Likewise, the choice of the link removed from $\gK$
is immaterial up to isomorphism, so $\gKm$
identifies a unique network.
The following expressions for individual expected benefits,
aggregate expected benefits, and welfare follow directly
from Lemma \ref{lem:cmpform} and \eqref{eq:welfare}.

\begin{proposition}
For every $n\ge3$, $p,\delta\in(0,1)$, and $c>0$,
the following expressions hold.

\begin{itemize}

\item \emph{Empty network} ($\gE$).
For every $i\in\N$,
\[
B_i(\gE)=0,\qquad
\Ben(\gE)=0,\qquad
\W(\gE)=0.
\]

\item \emph{Complete network} ($\gK$).
For every $i\in\N$,
\[
B_i(\gK)=(n-1)p^2\delta,
\]
\[
\Ben(\gK)=n(n-1)p^2\delta,
\qquad
\W(\gK)=n(n-1)(p^2\delta-c).
\]

\item \emph{Star network} ($\gS$).
For the center $h$ and each peripheral node $i\ne h$,
respectively,
\[
B_h(\gS)=(n-1)p^2\delta,
\qquad
B_i(\gS)=p^2\delta+(n-2)p^3\delta^2.
\]
The aggregate expected benefit and welfare are
\[
\Ben(\gS)
=(n-1)\bigl[2p^2\delta+(n-2)p^3\delta^2\bigr],
\]
\[
\W(\gS)
=(n-1)\bigl[2p^2\delta+(n-2)p^3\delta^2-2c\bigr].
\]

\item \emph{Complete network minus one link} ($\gKm$).
Let $u,v$ be the endpoints of the missing link.
Their individual expected benefits are
\[
B_u(\gKm)=B_v(\gKm)
=(n-2)p^2\delta
+p^2\delta^2(1-q^{n-2}),
\]
while for every $i\notin\{u,v\}$,
\[
B_i(\gKm)=(n-1)p^2\delta.
\]
The aggregate expected benefit and welfare are
\[
\Ben(\gKm)
=\bigl[n(n-1)-2\bigr]p^2\delta
+2p^2\delta^2(1-q^{n-2}),
\]
\[
\W(\gKm)
=\bigl[n(n-1)-2\bigr](p^2\delta-c)
+2p^2\delta^2(1-q^{n-2}).
\]

\item \emph{Cycle network} ($\gC$).
For $n=3$, the cycle coincides with the complete
network, so that for every $i\in\N$,
\[
B_i(\gC)=2p^2\delta,
\qquad
\Ben(\gC)=6p^2\delta,
\]
\[
\W(\gC)=6(p^2\delta-c).
\]
For $n=4$, the cycle is complete bipartite,
with two cells of size two. Hence, for every $i\in\N$,
\[
B_i(\gC)
=2p^2\delta+p^3(2-p)\delta^2,
\]
\[
\Ben(\gC)
=8p^2\delta+4p^3(2-p)\delta^2,
\]
\[
\W(\gC)
=8(p^2\delta-c)+4p^3(2-p)\delta^2.
\]

\end{itemize}
\end{proposition}

\rev{Table \ref{tab:notation} lists the network symbols, cost thresholds, and
stability notation that recur throughout the paper, including those introduced
in later sections.}

\begin{table}[htbp]
\centering
\caption{Network symbols, cost thresholds, and stability notation.}
\label{tab:notation}
\small
\begin{tabular}{@{}p{0.27\linewidth}p{0.68\linewidth}@{}}
\toprule
Symbol & Meaning \\
\midrule
$\gMP{a_1,\ldots,a_r}$ & Complete multipartite network with cell sizes $a_1,\ldots,a_r$ \\
$\gK$, $\gKm$, $\gE$, $\gS$, $\gC$ & Complete network, complete network minus one link, empty network, star, cycle \\
$g^p$ & Paw network, a triangle with one pendant link ($n=4$) \\
$\gA{k}$, $\gA{k,j}$ & Networks with $k$ links that are not complete multipartite and lie on the unrestricted benefit frontier ($n=5,6$) \\
$g_{[\rev{\eta}]}$ & Quotient network with cells of size $\rev{\eta}$ (Definition \ref{def:quotient}) \\
$c_{\mathrm{low}}$, $c_{\mathrm{mid}}$, $c_{\mathrm{high}}$ & Cost thresholds, defined in \eqref{eq:clow}, \eqref{eq:cmid}, and \eqref{eq:chigh} \\
$\mathcal I$ & Intermediate band of link counts, $n\le m\le\binom{n}{2}-1$ \\
$\PS(p,\delta,c)$ & Pairwise-stable networks \\
$\alpha(a)$, $\sigma(a)$ & Lower and upper ends of the stability window of $\gMP{a}$ \\
\bottomrule
\end{tabular}
\end{table}

\subsection{\texorpdfstring{\rev{The Optimization Problems and the Benefit Frontier}}{The Optimization Problems and the Benefit Frontier}}\label{sec:optimization_framework}

This \rev{subsection} introduces the optimization problems and the
benefit-frontier framework used throughout the analysis.
We distinguish between maximizing aggregate expected benefits
and maximizing welfare, which additionally accounts for \rev{link} costs.

We consider two classes of admissible networks. The
\emph{unrestricted problem} allows all networks in
$\G(\N)$, whereas the \emph{restricted problem} considers
only complete multipartite networks in
$\G^{\mathrm{mp}}(\N)$.
Both problems share the same objective but differ in their
admissible network classes.

To treat both problems within a common framework, let
$\mathcal A\in\{\G(\N),\G^{\mathrm{mp}}(\N)\}$
denote the admissible network class, and let
$\mathcal K_{\mathcal A}=\{|g|:g\in\mathcal A\}$
be the corresponding set of feasible link counts.
\rev{Since both problems are formulated in terms of networks,} the sets of benefit-optimal and
welfare-optimal solutions are understood as sets of
distinct networks, rather than graphs.

\begin{definition}[Optimality and benefit frontiers]
\label{def:optimality}
For each $k\in\mathcal K_{\mathcal A}$, define
\[
\Ben_{\mathcal A}^*(k)
=
\max_{\substack{g\in\mathcal A\\ |g|=k}}\Ben(g),
\qquad
\G_{\mathcal A}^*(k)
=
\argmax_{\substack{g\in\mathcal A\\ |g|=k}}\Ben(g).
\]
A network in $\G_{\mathcal A}^*(k)$ is
\emph{benefit-optimal} conditional on $k$ links.
The \emph{benefit frontier} is
\[
\mathcal F(\Ben_{\mathcal A}^*)
=
\{(k,\Ben_{\mathcal A}^*(k)):
k\in\mathcal K_{\mathcal A}\}.
\]
For fixed $n$, the benefit frontier depends on the survival
probability $p$ and the \rev{decay factor} $\delta$,
but not on the link cost $c$. \rev{Even} at a fixed link count, the benefit-maximizing networks may vary
with $p$ and $\delta$, and distinct networks may attain
the same maximum for a given parameter configuration.

For fixed $p$ and $\delta$, the \emph{upper concave
envelope} of the benefit frontier is the graph
of the smallest concave function on the convex
hull of $\mathcal K_{\mathcal A}$ that lies
weakly above all points of
$\mathcal F(\Ben_{\mathcal A}^*)$.
Equivalently, it is the upper boundary of the
convex hull of the benefit-frontier points.
The feasible link counts whose benefit-frontier
points lie on this envelope are called
\emph{supported link counts}. These include
both the vertices of the envelope and any
points lying along its linear segments.

The set of \emph{welfare-optimal networks} is
$\Eff_{\mathcal A}(p,\delta,c)
=\argmax_{g\in\mathcal A}\W(g)$.
The corresponding maximal welfare is
\[
W_{\mathcal A}^*(c)
=
\max_{g\in\mathcal A}\W(g)
=
\max_{k\in\mathcal K_{\mathcal A}}
\left\{\Ben_{\mathcal A}^*(k)-2ck\right\}.
\]
\end{definition}

For the unrestricted problem, we omit the subscript
$\mathcal A$ and write $\Ben^*(k)$, $\G^*(k)$,
$\mathcal F(\Ben^*)$, $\Eff(p,\delta,c)$,
and $W^*(c)$.
In this case,
$\mathcal K_{\mathcal A}
=\{0,\ldots,\binom{n}{2}\}$.
For the restricted problem, we use the superscript
$\mathrm{mp}$ and write $\Ben^{*,\mathrm{mp}}(k)$,
$\G^{*,\mathrm{mp}}(k)$,
$\mathcal F(\Ben^{*,\mathrm{mp}})$,
$\Eff^{\mathrm{mp}}(p,\delta,c)$, and
$W^{*,\mathrm{mp}}(c)$.
Since complete multipartite networks are identified
by integer partitions of $n$, the link counts attainable
within $\G^{\mathrm{mp}}(\N)$ need not be consecutive.

The benefit-frontier approach separates the identification
of benefit-optimal networks at each feasible link count
from the selection of the welfare-optimal link count.
While $\Ben_{\mathcal A}^*(k)$ determines the maximal
expected benefit attainable with $k$ links and
$\G_{\mathcal A}^*(k)$ identifies all networks attaining
it, welfare maximization selects among these link counts
by accounting for the cost $2ck$.

\rev{Adding a link never lowers the expected benefit of any node.}\footnote{This property also appears as Lemma 1 in \citet{LJG2025}.}

\begin{proposition}\label{prop:monotone}
Let $g\in\G(\N)$ and $ij\notin g$. Then
$B_m(g+ij)\ge B_m(g)$ for every $m\in\N$, with
strict inequality for $m\in\{i,j\}$. In particular,
$\Ben(g+ij)>\Ben(g)$.
\end{proposition}

\rev{Thus, link formation generates no negative externalities in terms of
connection benefits. In the unrestricted problem,
Proposition \ref{prop:monotone} implies that
$\Ben^*(k)$ is strictly increasing in $k$.
This argument does not directly extend to the
restricted problem, since adding a link need not
preserve the complete multipartite structure.}

Moreover, since
$\G^{\mathrm{mp}}(\N)\subseteq\G(\N)$, we have
$\Ben^{*,\mathrm{mp}}(k)\le\Ben^*(k)$
for every link count feasible in the restricted class,
and $W^{*,\mathrm{mp}}(c)\le W^*(c)$.
Equality in the latter holds if and only if
at least one unrestricted welfare-optimal network
is complete multipartite.

\rev{Potentially welfare-optimal link counts have a simple geometric characterization.}

\begin{proposition}\label{prop:hull}
Fix $n$, $p$, and $\delta$, and let
$\mathcal A\in\{\G(\N),\G^{\mathrm{mp}}(\N)\}$.
A feasible link count $k$ is attained by some
welfare-optimal network for some $c>0$ if and only if
$(k,\Ben_{\mathcal A}^*(k))$ lies on the upper
concave envelope of the benefit frontier.
\end{proposition}

In general, belonging to the upper concave
envelope of a benefit frontier is not sufficient
for a link count to be welfare-optimal at some
positive link cost, since the corresponding
supporting lines may have non-positive slopes.
The preceding characterization is therefore
specific to our model and follows from the
strict benefit monotonicity established in
Proposition~\ref{prop:monotone}.
In the analysis that follows, we mainly exploit
the exclusion implication of this characterization:
a benefit-frontier point lying strictly below
the upper concave envelope cannot correspond
to a welfare-optimal network at any positive
link cost.
It provides a geometric criterion
for excluding entire classes of networks from
welfare optimality, independently of the link
cost $c$, thereby simplifying the characterization
of welfare-optimal networks in both the unrestricted
and restricted problems.

Geometrically, welfare maximization selects the
points of the benefit frontier supported by a line
with slope $2c$. As $c$ increases, this slope
increases, shifting welfare-optimal link counts
weakly toward lower values. Points lying along
linear segments of the upper concave envelope
can also be welfare-optimal at the corresponding
switching costs.

\section{Welfare Optimality in the Unrestricted Network Space}\label{sec:efficiency}
We now study welfare maximization over the unrestricted
network space $\G(\N)$. Building on the benefit-frontier
framework introduced in Section \ref{sec:optimization_framework},
we characterize the structural properties of optimal
networks and identify the ranges of link costs
supporting their optimality.

\begin{revblock}
The analysis proceeds in five steps.
Section \ref{sec:perlinkbound} bounds the expected benefit per link.
Section \ref{sec:welfare} characterizes the two extremes of the cost range,
Section \ref{sec:star} the star and the intermediate band, and
Section \ref{sec:matchings} the role of matchings at the first threshold.
Section \ref{sec:n4} treats the case of four nodes completely.
Figure \ref{fig:regimes} summarizes our findings.
\end{revblock}

\subsection{\texorpdfstring{\rev{The Benefit-per-Link Bound}}{The Benefit-per-Link Bound}}\label{sec:perlinkbound}
We begin by establishing a global upper bound on
expected benefits per link.

\begin{theorem}[Benefit per link]\label{thm:perlink}
For every $g\in\G(\N)$ with $|g|\ge1$,
\begin{equation}\label{eq:perlink}
\frac{\Ben(g)}{|g|}
\le
2p^2\delta+(n-2)p^3\delta^2
=
\frac{\Ben(\gS)}{n-1}.
\end{equation}
Equality holds if and only if $g=\gS$.
\end{theorem}

\rev{\conf{The proof bounds, in each survival state, the benefit of the residual network by its number of links times $2\delta+(|S|-2)\delta^2$, the benefit per link of a star spanning the surviving nodes. Only a star attains this bound in every state.}}
\redtxt{Since each link costs $2c$ in aggregate, \eqref{eq:perlink} also gives the largest expected benefit per unit of aggregate link cost.}
This global bound will be useful for
characterizing welfare-optimal networks
when \rev{link} costs are high.

\subsection{\texorpdfstring{\rev{Extreme Link Costs}}{Extreme Link Costs}}\label{sec:welfare}
We now turn to welfare optimality and study how the optimal
network changes when the \rev{link cost} is either
sufficiently low or sufficiently high.
The two extremes of the cost range admit a complete
characterization.
\begin{theorem}\label{thm:extreme}
Define
\begin{revblock}
\begin{align}
c_{\mathrm{low}}
&=
p^2\left[\delta-\delta^2+q^{\,n-2}\delta^2\right],
\label{eq:clow}\\
c_{\mathrm{high}}
&=
p^2\delta+\frac{n-2}{2}p^3\delta^2.
\label{eq:chigh}
\end{align}
\end{revblock}
\rev{Then the following statements hold.}

\begin{enumerate}[label=(\roman*),leftmargin=2.2em]
\item
The complete network satisfies
$\gK\in\Eff(p,\delta,c)$ if and only if
$c\leq c_{\mathrm{low}}$, and
$\Eff(p,\delta,c)=\{\gK\}$ if and only if
$c<c_{\mathrm{low}}$.
At $c=c_{\mathrm{low}}$,
$\gK,\gKm\in\Eff(p,\delta,c_{\mathrm{low}})$.
\item
The empty network satisfies
$\gE\in\Eff(p,\delta,c)$ if and only if
$c\geq c_{\mathrm{high}}$, and
$\Eff(p,\delta,c)=\{\gE\}$ if and only if
$c>c_{\mathrm{high}}$.
At $c=c_{\mathrm{high}}$,
$\Eff(p,\delta,c)=
\{\gE, \gS\}$.
\end{enumerate}
\end{theorem}

Theorem~\ref{thm:extreme} thus gives exact thresholds at both extremes of the cost range.\footnote{The threshold $c_{\mathrm{low}}$
coincides with the sufficient condition
established by \citet{LJG2025} for the
optimality of the complete network.
Part~(i) additionally establishes necessity.
For the empty network, part~(ii) provides
an exact characterization, sharpening the
sufficient condition in \citet{LJG2025}.}

\subsection{\texorpdfstring{\rev{The Star and the Intermediate Band}}{The Star and the Intermediate Band}}\label{sec:star}
We now turn to intermediate costs and show that the star is uniquely optimal over a nonempty interval immediately below $c_{\mathrm{high}}$.
\begin{theorem}\label{thm:star}
There exists a threshold $c_{\mathrm{mid}}<c_{\mathrm{high}}$ such that
$
\Eff(p,\delta,c)=\{\gS\}$
for all $c\in(c_{\mathrm{mid}},c_{\mathrm{high}}),$
where
\begin{revblock}
\begin{equation}\label{eq:cmid}
c_{\mathrm{mid}}
=
\max_{n\le k\le\binom{n}{2}}
\frac{\Ben^{*}(k)-\Ben(\gS)}
     {2\bigl(k-(n-1)\bigr)}.
\end{equation}
\end{revblock}
At $c=c_{\mathrm{mid}}$, the star remains welfare-optimal and is co-optimal
with at least one network having more than $n-1$ links.
\end{theorem}
Theorem~\ref{thm:star} identifies the exact lower
threshold $c_{\mathrm{mid}}$ of the star's welfare-optimality
interval.
This threshold is constructed by comparing the star
with networks having more than $n-1$ links.
The key step relies on the benefit-per-link bound
established in Theorem~\ref{thm:perlink}, which rules
out networks with fewer than $n-1$ links whenever
$c<c_{\mathrm{high}}$ and ensures that
$c_{\mathrm{mid}}<c_{\mathrm{high}}$.
\footnote{\citet{LJG2025} establish a sufficient
condition for the welfare optimality of the star over
a cost interval with only upper endpoint $c_{\mathrm{high}}$.}

\rev{The thresholds $c_{\mathrm{low}}$, $c_{\mathrm{mid}}$, and $c_{\mathrm{high}}$ govern the
transitions between welfare-optimal networks as the link cost increases.
They are ordered as follows. In particular, a non-degenerate cost interval
separates the optimality regions of the complete network and the star
whenever $n\ge4$.}

\begin{proposition}\label{prop:thresholds}
The three thresholds satisfy
$c_{\mathrm{low}}\leq c_{\mathrm{mid}}<c_{\mathrm{high}}.$
Moreover, $c_{\mathrm{low}}=c_{\mathrm{mid}}$ if $n=3$, whereas
$c_{\mathrm{low}}<c_{\mathrm{mid}}$ if $n>3$.
\end{proposition}

\begin{figure}[htbp]
\centering
\begin{tikzpicture}[x=1cm,y=1cm,>=latex]
\draw[->,thick] (0,0) -- (13.4,0) node[right] {$c$};
\foreach \x/\l in {0/$0$,3.3/$c_{\mathrm{low}}$,7.5/$c_{\mathrm{mid}}$,10.5/$c_{\mathrm{high}}$}
  \draw (\x,0.12) -- (\x,-0.12) node[below] {\l};
\node[above,align=center] at (1.65,0.2) {complete\\$\gK$};
\node[above,align=center] at (5.4,0.2) {intermediate networks\\$n\le|g|\le\binom{n}{2}-1$};
\node[above,align=center] at (9.0,0.2) {star\\$\gS$};
\node[above,align=center] at (11.95,0.2) {empty\\$\gE$};
\node[below,align=center,font=\footnotesize] at (3.3,-0.75) {$\gK$ minus\\any matching};
\node[below,align=center,font=\footnotesize] at (7.5,-0.75) {$\gS$ and a\\denser network};
\node[below,align=center,font=\footnotesize] at (10.5,-0.75) {$\gS$ and $\gE$};
\end{tikzpicture}
\caption{\rev{Welfare-optimal networks as the link cost $c$ varies, for $n\ge4$.
Labels above the axis give the optimal networks on each open interval. Labels below the axis
give the networks that are co-optimal at each threshold. The thresholds are defined
in \eqref{eq:clow}, \eqref{eq:cmid}, and \eqref{eq:chigh}. For $n=3$,
$c_{\mathrm{low}}=c_{\mathrm{mid}}$ and the intermediate regime is empty.}}
\label{fig:regimes}
\end{figure}
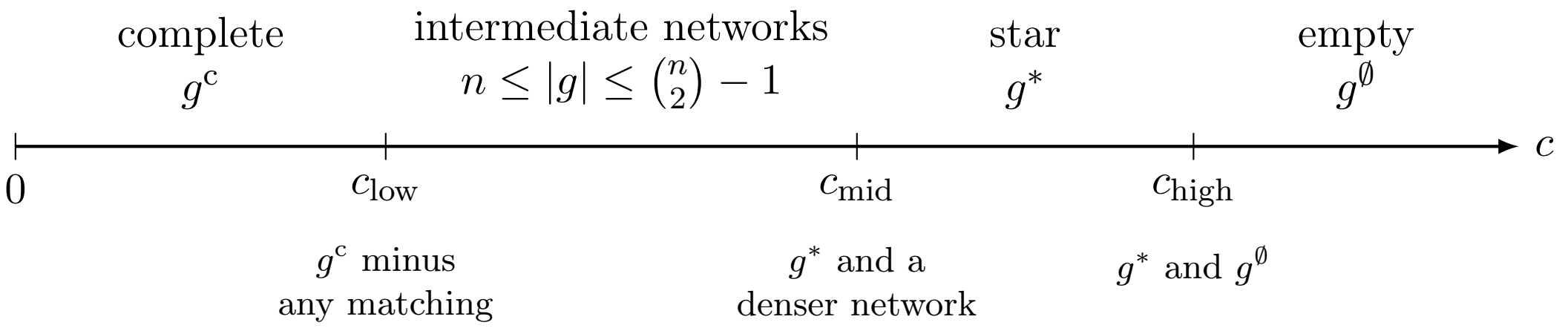

\redtxt{Figure \ref{fig:regimes} summarizes the resulting sequence of optimal networks as the link cost increases.}

\rev{Some networks can be ruled out on the basis of their number of links alone,
whatever their structure and the link cost.}

\begin{corollary}\label{cor:linkcounts}
No network with $1\leq |g|\leq n-2$ links
is welfare-optimal for any $c\rev{>0}$.
\end{corollary}

We define the intermediate band as the set of link counts strictly between those
of the star and the complete \rev{network, that is,}
\[
\mathcal{I}
=
\left\{
m\in\mathbb{N}:
n\leq m\leq \binom{n}{2}-1
\right\}.
\]
Whether a link count $m\in\mathcal{I}$ supports a welfare-optimal network depends on the frontier and the link cost. For $n=3$, the intermediate band is empty, and the welfare-optimal networks are already completely characterized by the preceding results, since $c_{\mathrm{low}}=c_{\mathrm{mid}}$.

The analysis of intermediate networks is therefore relevant only for $n\geq4$.
\begin{corollary}\label{cor:intermediate}
Let $n\geq 4$. For every
$c\in(c_{\mathrm{low}},c_{\mathrm{mid}})$,
every welfare-optimal network $g$ satisfies
$|g|\in\mathcal{I}$.
\end{corollary}

\subsection{\texorpdfstring{\rev{Matchings and the First Threshold}}{Matchings and the First Threshold}}\label{sec:matchings}
A matching of a network $g\in\G(\N)$
is a set of links in a graph representative
of $g$ such that no two distinct links
share an endpoint.
The empty set is also a matching.
A maximum matching is a matching
of maximum cardinality.
In what follows, we focus on matchings of the complete
network $\gK=K_{\N}$. Every maximum matching of $\gK$
has cardinality $\lfloor n/2\rfloor$.
Removing a matching $M$ from $\gK$ preserves a particular
connectivity property\rev{. For} every missing link $ij\in M$,
the endpoints $i$ and $j$ remain connected through any
other surviving node. Thus, whenever $i$ and $j$ survive
along with at least one additional node, their distance
in the residual network is exactly two.
\rev{\conf{Removing a matching is therefore the cheapest way, in terms of forgone
benefits, to delete a given number of links from the complete network.}}

\begin{theorem}\label{prop:matching}
Let $\chi=\lfloor n/2\rfloor$.
At $c=c_{\mathrm{low}}$, the welfare-optimal
networks are exactly those obtained from
$\gK$ by removing a matching, including
the empty \rev{matching, that is,}
\[
\Eff(p,\delta,c_{\mathrm{low}})
=
\left\{
\gK\setminus M:
M\text{ is a matching of }\gK
\right\}.
\]
In particular,
\[
|\Eff(p,\delta,c_{\mathrm{low}})|=\chi+1.
\]
\end{theorem}

These networks admit multiple graphs. The number of matchings containing $\rev{\rho}$ links is
$
\binom{n}{2\rev{\rho}}\frac{(2\rev{\rho})!}{2^{\rev{\rho}} \rev{\rho}!}
=
\frac{n!}{2^{\rev{\rho}} \rev{\rho}!(n-2\rev{\rho})!}.
$
Hence, the total number of graphs is
$$
\sum_{\rev{\rho}=0}^{\chi}
\frac{n!}{2^{\rev{\rho}} \rev{\rho}!(n-2\rev{\rho})!}.
$$
\rev{Only one of these co-optimal networks remains welfare-optimal when the link
cost rises slightly above $c_{\mathrm{low}}$.}

\begin{theorem}\label{prop:right_low}
Let $\chi=\lfloor n/2\rfloor$.
There exists $\varepsilon>0$ such that,
for every $c\in(c_{\mathrm{low}},c_{\mathrm{low}}+\varepsilon)$,
the unique welfare-optimal network is represented
by \rev{a graph} obtained from $\gK$ by removing a
maximum matching.
\end{theorem}

The number of distinct labeled graphs
representing this network is
\[
\frac{n!}{2^\chi\chi!(n-2\chi)!}.
\]
\redtxt{For $n\geq4$, all networks obtained from $\gK$ by removing nonempty matchings belong to the intermediate band $\mathcal I$.}

\rev{By contrast, removing links that share a node is never optimal.}

\begin{theorem}\label{prop:never_optimal}
Let $n\geq4$ and
$2\leq \rev{\rho}\leq\lfloor n/2\rfloor$.
Every network obtained from $\gK$ by removing
$\rev{\rho}$ links that do not form a matching
is never welfare-optimal, regardless of the
link cost.
\end{theorem}

For $n\geq4$, all networks considered in Theorem \ref{prop:never_optimal} have a number of links belonging to the intermediate band $\mathcal I$, since
\[
|g|=\binom n2-\rev{\rho}\in\mathcal I
\qquad
\text{for }
2\leq \rev{\rho}\leq\lfloor n/2\rfloor.
\]
\rev{Membership in the intermediate band therefore does not guarantee welfare
optimality. Whether a network is welfare-optimal depends on its structure as
well as on its number of links.}

\subsection{\texorpdfstring{\rev{The Case of Four Nodes}}{The Case of Four Nodes}}\label{sec:n4}
\rev{The simplest nontrivial case is $n=4$, for which the intermediate band is
$\mathcal{I}=\{4,5\}$. Here the welfare-optimal networks can be characterized
completely, including the switching costs at which several networks are co-optimal.}

\begin{proposition}\label{prop:n4}
Let $n=4$. The \rev{welfare-optimal networks are given by}
\[
\Eff(p,\delta,c)=
\begin{cases}
\{\gK\},
& 0<c<c_{\mathrm{low}},\\[2mm]
\{\gK,\gKm,\gC\},
& c=c_{\mathrm{low}},\\[2mm]
\{\gC\},
& c_{\mathrm{low}}<c<c_{\mathrm{mid}},\\[2mm]
\{\gC, \gS\},
& c=c_{\mathrm{mid}},\\[2mm]
\{\gS\},
& c_{\mathrm{mid}}<c<c_{\mathrm{high}},\\[2mm]
\{\gS, \gE\},
& c=c_{\mathrm{high}},\\[2mm]
\{\gE\},
& c>c_{\mathrm{high}}.
\end{cases}
\]
In particular, the cycle $\gC$ is the unique \rev{welfare-optimal}
network throughout the non-degenerate interval
$c_{\mathrm{low}}<c<c_{\mathrm{mid}},$
with
$c_{\mathrm{mid}}-c_{\mathrm{low}}
=3p^3\rev{q}\delta^2>0.$
Moreover,
$c_{\mathrm{high}}-c_{\mathrm{mid}}=2p^4\delta^2>0.$
\end{proposition}

\begin{figure}[htbp]
\centering
\begin{tikzpicture}[x=1.5cm,y=1.4cm,>=latex]
\draw[->] (0,0) -- (6.8,0) node[right] {$k$};
\draw[->] (0,0) -- (0,4.0) node[above] {$\Ben^*(k)$};
\foreach \x in {0,...,6} \draw (\x,0.05) -- (\x,-0.05) node[below] {\small$\x$};
\foreach \y in {1,2,3} \draw (0.05,\y) -- (-0.05,\y) node[left] {\small$\y$};
\draw[thick] (0,0) -- (3,2.557) -- (4,3.078) -- (6,3.456);
\foreach \pt in {(0,0),(3,2.557),(4,3.078),(5,3.267),(6,3.456)} \fill \pt circle (2pt);
\foreach \pt in {(1,0.576),(2,1.428),(4,2.857)} \draw[fill=white] \pt circle (2pt);
\node[anchor=west] at (0.2,0.12) {$\gE$};
\node[above left] at (3,2.557) {$\gS$};
\node[above] at (4,3.12) {$\gC$};
\node[above] at (5,3.31) {$\gKm$};
\node[above right] at (6,3.456) {$\gK$};
\node[anchor=west] at (4.08,2.857) {$g^p$};
\node[anchor=east,font=\small] at (1.35,1.5) {slope $2c_{\mathrm{high}}$};
\node[anchor=north west,font=\small] at (3.05,2.42) {slope $2c_{\mathrm{mid}}$};
\node[anchor=north west,font=\small] at (5.1,3.2) {slope $2c_{\mathrm{low}}$};
\end{tikzpicture}
\caption{\rev{Benefit frontier for $n=4$, drawn for $p=0.6$ and $\delta=0.8$, with the
number of links $k$ on the horizontal axis. Filled points lie on the upper concave
envelope (solid line), whose slopes are twice the thresholds. The hollow points at
one and two links are frontier points below the envelope. The paw network $g^p$
lies below the frontier at four links.}}
\label{fig:n4frontier}
\end{figure}
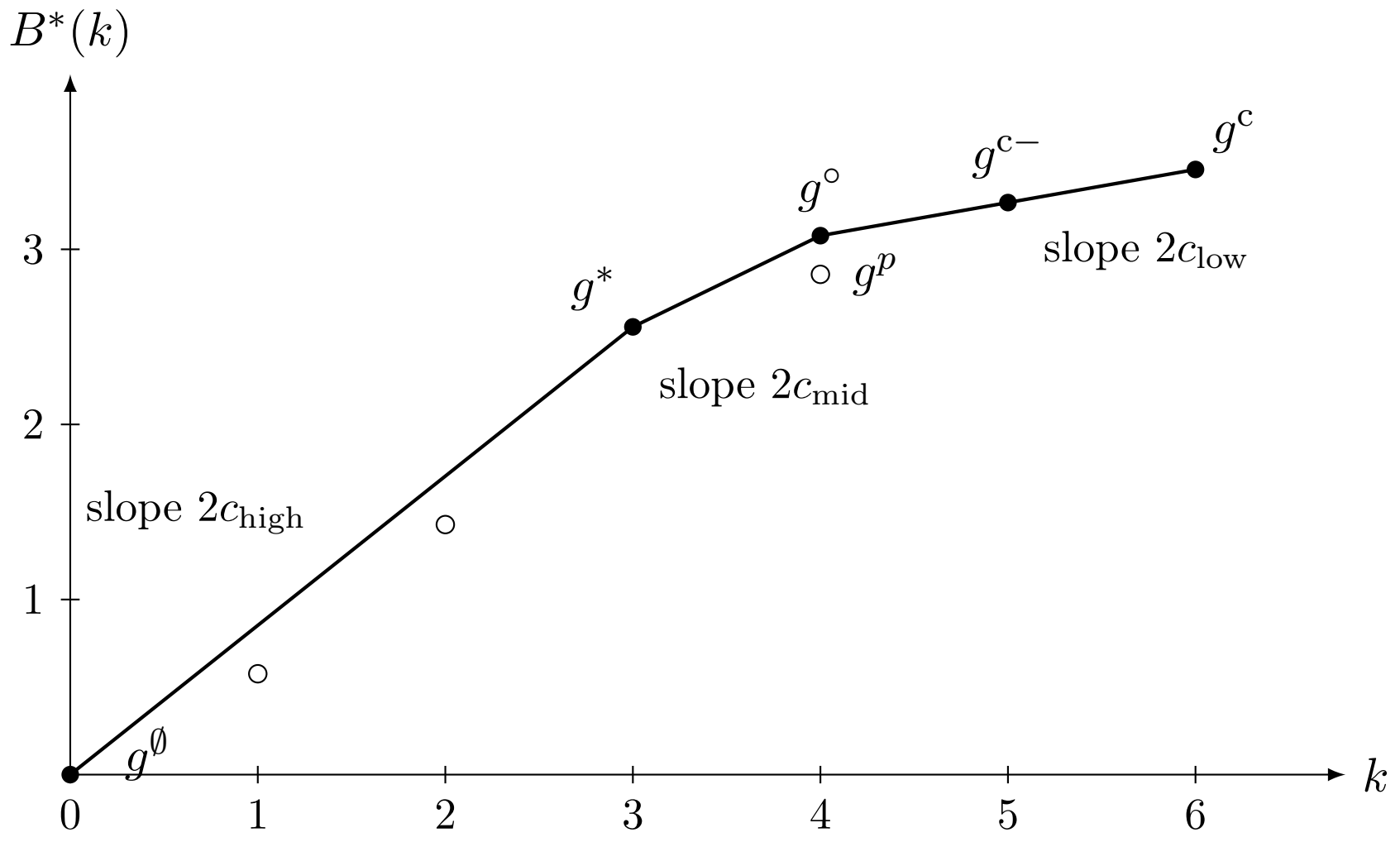

\redtxt{Figure \ref{fig:n4frontier} shows the frontier behind Proposition \ref{prop:n4}.}

\rev{Belonging to the intermediate band does not guarantee welfare optimality. The} five-link network $\gKm$ is welfare-optimal only at $c=c_{\mathrm{low}}$, while the four-link paw network $g^p$\rev{, a triangle with one pendant link,} is never welfare-optimal. This distinction also illustrates the role of matchings. Removing two links that do not share a node from $\gK$ produces the cycle $\gC$, whereas removing two links that share a node produces the paw network $g^p$. Only the former corresponds to removing a matching, consistently with Theorems \ref{prop:matching} and \ref{prop:never_optimal}. 

\rev{The nature of co-optimality also differs across the three thresholds.} At $c=c_{\mathrm{low}}$, co-optimality is a structural feature of the model\rev{. For} any admissible values of $p$ and $\delta$, removing any matching from the complete network preserves welfare optimality, giving rise to an increasing number of co-optimal networks as $n$ grows. At $c=c_{\mathrm{high}}$, by contrast, only the star and the empty network are co-optimal, up to isomorphism, regardless of $n$, $p$, and $\delta$. The situation at $c=c_{\mathrm{mid}}$ is more intricate. Although the star is necessarily co-optimal with at least one denser network, additional co-optimal networks may emerge for particular combinations of \rev{the parameters} $p$ and $\delta$.

Proposition~\ref{prop:n4} illustrates the sufficiency
direction of Proposition~\ref{prop:hull}.
Although the characterization for $n=4$ could
also be obtained through direct comparisons
of the welfare functions of the relevant networks,
the geometric approach illustrates how the
upper concave envelope can be used to identify
the exact switching costs and the complete
correspondence of welfare-optimal networks.
Elsewhere in the paper, we primarily exploit
the necessity direction of Proposition~\ref{prop:hull}
to exclude networks from welfare optimality.

\section{Welfare Optimality in the Restricted Network Space}\label{sec:computation}

\rev{The preceding section establishes} several properties of
welfare-optimal networks in the unrestricted class.
However, characterizing the unrestricted benefit frontier
becomes increasingly difficult as the number of nodes grows,
even for relatively small populations.
\rev{We therefore turn to the restricted problem.
Section \ref{sec:smalln} examines what the restriction to complete multipartite
networks costs in small networks, Section \ref{sec:algorithm} develops the algorithm,
Section \ref{sec:misspec} studies the cost of misjudging failure risk, and Section \ref{sec:heuristics} compares the exact approach
with heuristics.}

\subsection{\texorpdfstring{\rev{What the Restriction Costs in Small Networks}}{What the Restriction Costs in Small Networks}}\label{sec:smalln}
We examine the cases $n=4,5,6$ to identify conditions
under which restricting attention to complete multipartite
networks entails no loss of welfare optimality.
\rev{For $n=4$ there is no loss for any parameter values. For $n=5$ and $n=6$ we
give sufficient conditions on $p$ and $\delta$.}
By Corollary \ref{cor:linkcounts}, networks with
$1\leq |g|\leq n-2$ cannot be welfare-optimal, while
Theorem \ref{thm:perlink} implies that only stars can be
welfare-optimal among networks with exactly $n-1$ links.
Since the empty network, stars, and the complete network
are complete multipartite, any discrepancy between the
restricted and unrestricted optimal correspondences must
arise within the intermediate \rev{band $\mathcal I$.}
\begin{proposition}\label{prop:n4mp}
Let $n=4$. Then, for every $p,\delta\in(0,1)$ and $c>0$,
$\Eff(p,\delta,c)=\Eff^{\mathrm{mp}}(p,\delta,c)$.
\end{proposition}

For $n=5$, the complete multipartite class first
leaves a gap at a link count relevant for welfare
optimality.
To examine this and subsequent cases, we introduce
an exclusion criterion based on the geometric
characterization established in
Proposition \ref{prop:hull}.
\begin{proposition}\label{prop:chord_exclusion}
Let $g^-$, $g$, and $g^+$ be three networks such that
$m^-:=|g^-|<|g|=:m<|g^+|=:m^+$,
and define
$\theta=\frac{m^+-m}{m^+-m^-}$.
If
\begin{equation}\label{eq:chord_exclusion}
\Ben(g)
<
\theta\Ben(g^-)+(1-\theta)\Ben(g^+),
\end{equation}
then $g$ cannot be welfare-optimal for any $c>0$.
\end{proposition}

Geometrically, condition \eqref{eq:chord_exclusion}
means that $(m,\Ben(g))$ lies strictly below the
chord joining $(m^-,\Ben(g^-))$ and
$(m^+,\Ben(g^+))$ in the \rev{plane of link counts and benefits}.
Neither $g^-$ nor $g^+$ needs to be welfare-optimal.
The converse does not hold, since a network lying
above one chord may still lie below another chord,
or more generally below the unrestricted upper
concave envelope.

\begin{proposition}\label{prop:n5}
Let $n=5$. If $p(1+10\delta)<5$, then
$\Eff(p,\delta,c)=\Eff^{\mathrm{mp}}(p,\delta,c)$
for every $c>0$.\footnote{\rev{It can be shown that this sufficient condition is also necessary.}}
\end{proposition}

\rev{\conf{The condition $p(1+10\delta)<5$ holds for every $\delta$ when $p\le5/11$, and it can
fail only if $p>5/11$ and $\delta>2/5$, that is, when nodes are fairly reliable and
benefits decay slowly.} The excluded region covers about one fifth of the parameter
square (Table \ref{tab:mp_comparison}). \conf{When the condition fails, the only network
outside the complete multipartite class that may be welfare-optimal is the five-cycle.}}

We next consider $n=6$.

\begin{proposition}\label{prop:n6}
Let $n=6$, and define
$
Q(p,\delta)
=
(9\delta+2)p^2-(27\delta+6)p+12.
$
If $Q(p,\delta)>0$, then
$\Eff(p,\delta,c)=\Eff^{\mathrm{mp}}(p,\delta,c)$
for every $c>0$.\footnote{\rev{It can be shown that this sufficient condition is also necessary.}}
Equivalently, \rev{$Q(p,\delta)>0$ if and only if}
\[
\delta<
\frac{2(p^2-3p+6)}{9p(3-p)}.
\]
\end{proposition}

\rev{\conf{As for $n=5$, the condition $Q(p,\delta)>0$ can fail only when both parameters are
large. It holds for every $\delta$ when $p\le(33-\sqrt{561})/22\approx0.42$, and for
every $p$ when $\delta\le4/9$.}}

\rev{We do not extend the comparison beyond six nodes. Already for $n=7$ there are} $1{,}044$
non-isomorphic networks,
of which only $15$ are complete multipartite\rev{, and identifying} the unrestricted benefit frontier and
determining which networks can be welfare-optimal
\rev{becomes} increasingly demanding.
\conf{For given parameter values, however, the comparison can be carried out
numerically. At $p=1/10$ and $\delta=1/2$, exhaustive search shows that the best
complete multipartite network remains optimal among all networks at every link
cost for $n=7$, $8$, and $9$ (Section \ref{sec:heuristics}).}
Table \ref{tab:mp_comparison} compares the reduction
in admissible networks with the potential loss
of exactness in welfare optimization \rev{for $n=4,5,6$.}
The results show that eliminating most networks
need not entail a comparable loss of generality.

\begin{table}[htbp]
\centering
\caption{Reduction in admissible networks and potential
loss of exactness in welfare optimization.}
\label{tab:mp_comparison}
\begin{tabular}{@{}crrrr@{}}
\toprule
$n$
& \shortstack{Unrestricted\\networks}
& \shortstack{Complete multipartite\\networks}
& \shortstack{\rev{Network}\\reduction (\%)}
& \shortstack{Potential loss\\of exactness\\(at most \%)} \\
\midrule
$4$ & $11$  & $5$  & $54.55$ & $0.00$  \\
$5$ & $34$  & $7$  & $79.41$ & $20.58$ \\
$6$ & $156$ & $11$ & $92.95$ & $21.02$ \\
\bottomrule
\end{tabular}

\medskip
\begin{minipage}{0.95\textwidth}
\footnotesize
\emph{Notes:}
All networks are counted up to graph isomorphism.
Network reduction is the percentage of unrestricted
networks excluded by restricting attention to
complete multipartite networks.
The last column reports an upper bound on the
measure, expressed as a percentage of the area $(0,1)^2$,
of the region in which the restricted and unrestricted
welfare-optimal network sets may differ for some $c>0$.
The bounds follow from the sufficient conditions
established above.
\end{minipage}
\end{table}
This \rev{table} motivates restricting attention to complete
multipartite networks. Their representation through
integer partitions of $n$ substantially reduces the
optimization problem and provides the basis for the
dynamic programming approach developed next.

\subsection{The \texorpdfstring{\rev{Algorithm}}{Algorithm}}\label{sec:algorithm}

We now turn to the computation of the restricted benefit frontier
within the class of complete multipartite networks.
Using their representation through integer partitions and the
closed-form expression for expected benefits established in
Lemma \ref{lem:cmpform}, we develop exact procedures for
identifying welfare-optimal networks.
We first present an enumeration algorithm and then exploit
the additive structure of expected benefits to obtain a
polynomial-time dynamic program. \rev{We subsequently examine the cost of
misjudging failure risk and compare the exact approach with heuristic
procedures.} The enumeration procedure is
summarized in Algorithm \ref{alg:frontier}.

\begin{algorithm}[htbp]
\caption{Frontier under node failure risk}\label{alg:frontier}
\begin{algorithmic}[1]
\Require $n$, $p$, $\delta$
\Ensure the restricted benefit frontier,
the envelope vertices, and the switching costs
\State $\mathcal{B}\gets\emptyset$
  \Comment{Best benefit found at each link count}
\ForAll{partitions $a=(a_1,\dots,a_r)$ of $n$}
  \State $k\gets\tfrac12\bigl(n^{2}-\sum_t a_t^{2}\bigr)$
    \Comment{$|g|$, by \eqref{eq:cmpL}}
  \State $B\gets 2p^{2}\delta k+p^{2}\delta^{2}
          \sum_t a_t(a_t-1)\bigl(1-\rev{q^{\,n-a_t}}\bigr)$
    \Comment{by \eqref{eq:cmpB}}
  \If{$k\notin\mathcal{B}$ \textbf{or} $B>\mathcal{B}[k].\mathit{benefit}$}
    \State $\mathcal{B}[k]\gets(B,a)$
  \EndIf
\EndFor
\State $H\gets$ vertices of the upper concave envelope of
  $\{(k,\mathcal{B}[k].\mathit{benefit})\}$
\State \Return $\mathcal{B}$, the networks at the vertices of $H$ in decreasing order of
  $k$, and the \rev{switching costs} $c=\tfrac{\Delta B}{2\Delta k}$ along
  consecutive edges of $H$
\end{algorithmic}
\end{algorithm}
The stored frontier also includes any nonvertex points that become welfare-optimal at switching costs.
The enumeration procedure is exact, but its computational cost grows with the number of partitions of $n$.

\begin{proposition}[Computational complexity]\label{prop:complexity}
Algorithm \ref{alg:frontier} returns the exact restricted frontier using
$O\bigl(n\,\mathrm{p}(n)+n^{2}\bigr)$
arithmetic operations, where $\mathrm{p}(n)$ denotes the number of integer partitions of $n$. By the \rev{asymptotic formula of Hardy and Ramanujan,}
$
\mathrm{p}(n)
\sim
\frac{\exp\!\bigl(\pi\sqrt{2n/3}\bigr)}
     {4n\sqrt{3}},
$
the computational complexity is
$
\exp\bigl(\Theta(\sqrt n)\bigr).
$
\end{proposition}

The number of non-isomorphic networks grows as
$\exp(\Theta(n^2))$, whereas the number of
complete multipartite networks grows as
$\exp(\Theta(\sqrt n))$.
Although this is a substantial reduction,
the additive structure of \eqref{eq:cmpB} allows us
to compute the restricted frontier in polynomial time,
without enumerating the partitions.

For fixed $n$ and $q=1-p$, define the function
$f:\{1,\ldots,n\}\to\mathbb{R}$ by
$f(m)=m(m-1)\bigl(1-q^{n-m}\bigr).$
For any partition $a=(a_1,\ldots,a_r)\vdash n$,
let
$s(a)=\sum_t a_t^2.$
By Lemma \ref{lem:cmpform},
\[
|\gMP{a}|=\frac12\bigl(n^2-s(a)\bigr),
\qquad
\Ben(\gMP{a})
=
p^2\delta\bigl(n^2-s(a)\bigr)
+
p^2\delta^2\sum_t f(a_t).
\]
Since $|\gMP{a}|=(n^2-s(a))/2$, fixing $s(a)=s$ is
equivalent to fixing the number of links.
Moreover, in the expression for $\Ben(\gMP{a})$,
the term $p^2\delta(n^2-s)$ is constant across
all partitions satisfying $s(a)=s$.
Hence, maximizing the expected benefit conditional
on a given link count reduces to maximizing
\[
\sum_t f(a_t)
\]
over all partitions satisfying $\sum_t a_t=n,$
and  $\sum_t a_t^2=s.$

The additivity of both constraints and the objective
leads to the following dynamic program.
\begin{theorem}[Polynomial computation of the restricted frontier]
\label{thm:dp}
For $j\in\{0,\ldots,n\}$ and $s\in\{0,\ldots,n^{2}\}$, define
\[
\Phi(j,s)
=
\max\left\{
\sum_t f(a_t):
\sum_t a_t=j,\;
\sum_t a_t^{2}=s
\right\},
\]
where the maximum is taken over multisets of positive cell sizes. Set
$\Phi(0,0)=0$
and $\Phi(j,s)=-\infty$ whenever no such multiset exists. For $1\le j\le n$ and $0\le s\le n^2$, $\Phi$
satisfies the recursion
\begin{equation}
\label{eq:dprec}
\Phi(j,s)
=
\max_{1\le \rev{m}\le\min\{j,\lfloor\sqrt{s}\rfloor\}}
\left\{
\Phi(j-\rev{m},s-\rev{m}^{2})+f(\rev{m})
\right\}.
\end{equation}
For each feasible $s$, the maximum benefit among complete multipartite
networks with
$|\gMP{a}|=\frac{1}{2}(n^{2}-s)$
links is therefore
\begin{equation}
\label{eq:dpfrontier}
\Ben^{*,\mathrm{mp}}\!\left(\frac{1}{2}(n^{2}-s)\right)
=
p^{2}\delta(n^{2}-s)
+
p^{2}\delta^{2}\Phi(n,s).
\end{equation}
The recursion can be evaluated over all states using
$O(n^{4})$ arithmetic operations and $O(n^{3})$ space.
\end{theorem}

\rev{Thus, although the number of complete multipartite networks grows
superpolynomially with $n$, both the maximum welfare and a corresponding
welfare-optimal network can be computed, for any given $p$, $\delta$, and $c$,
in $O(n^4)$ arithmetic operations.}

\rev{In the numerical implementation, the dynamic program also stores, for each feasible value of $s$, a partition that maximizes benefits.} Once these partitions are recovered, their benefits
can be evaluated directly from \eqref{eq:cmpB} before
constructing the upper concave envelope.

Table \ref{tab:scaling} illustrates the computational
performance of the dynamic program for selected values of $n$. For comparison, the table also reports \rev{the number of labeled graphs on $n$ nodes} and the number of integer partitions of $n$. The last column reports the number of vertices of the upper concave envelope. We focus on networks associated with
these vertices, as they remain optimal over non-degenerate intervals of link costs.

\begin{table}[htbp]
\centering
\caption{Cost of computing the restricted benefit frontier at
$p=1/10$ and $\delta=1/2$ using the dynamic program of
Theorem \ref{thm:dp}. The search is performed in floating-point
arithmetic, while every reported network is recomputed and
compared using exact rational arithmetic. \rev{At $n=300$ double precision
no longer separates competing networks, whose benefits differ by terms of
order $q^{n}$, and the number of envelope vertices in that row is computed
in exact integer arithmetic.} The empirical growth
exponent is $3.9$, consistent with the
$O(n^4)$ bound.}
\label{tab:scaling}
\begin{tabular}{@{}rrrrr@{}}
\toprule
$n$
& \rev{labeled graphs} $2^{\binom{n}{2}}$
& partitions $\mathrm{p}(n)$
& time (s)
& envelope vertices \\
\midrule
$30$  & $\approx 10^{130}$   & $5{,}604$            & $0.01$  & $25$  \\
$50$  & $\approx 10^{368}$   & $204{,}226$          & $0.04$  & $42$  \\
$70$  & $\approx 10^{726}$   & $4{,}087{,}968$      & $0.18$  & $59$  \\
$100$ & $\approx 10^{1490}$  & $190{,}569{,}292$    & $0.27$  & $84$  \\
$150$ & $\approx 10^{3364}$  & $4.1\times 10^{10}$  & $1.11$  & $124$ \\
$200$ & $\approx 10^{5990}$  & $4.0\times 10^{12}$  & $2.88$  & $166$ \\
$300$ & $\approx 10^{13501}$ & $9.3\times 10^{15}$  & $14.39$ & \rev{$251$} \\
\bottomrule
\end{tabular}
\end{table}

\begin{figure}[htbp]
\centering
\includegraphics[width=\textwidth]{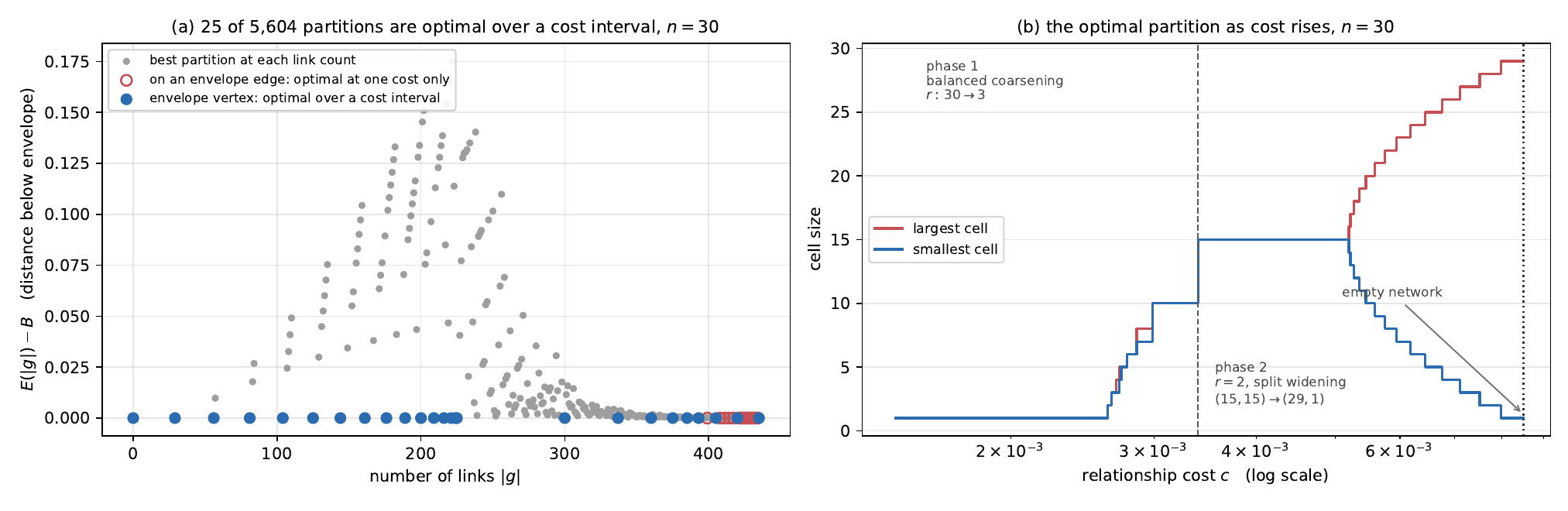}
\caption{Restricted benefit frontier at $n=30$, $p=1/10$, and
$\delta=1/2$. Panel (a) reports the distance between the maximum
benefit at each attainable link count and the upper concave envelope.
Panel (b) reports the largest and smallest cell sizes along the
sequence of networks associated with the envelope vertices.}
\label{fig:frontier30}
\end{figure}

We next illustrate the structure of the frontier for
$n=30$, $p=1/10$, and $\delta=1/2$. For this parameterization,
the restricted upper concave envelope has twenty-five vertices.
Table \ref{tab:frontier30} reports selected networks
associated with these vertices, while Figure \ref{fig:frontier30}
displays the complete frontier and the corresponding evolution
of cell sizes.

\begin{table}[htbp]
\centering
\caption{Selected networks associated with the upper-concave-envelope
vertices at $n=30$, $p=1/10$, and $\delta=1/2$, ordered by increasing
link cost.}
\label{tab:frontier30}
\begin{tabular}{@{}rrl@{}}
\toprule
$|g|$ & cells $r$ & network \\
\midrule
$435$ & $30$ & $\gMP{1,\ldots,1}$ \quad (complete network) \\
$420$ & $15$ & $\gMP{2,\ldots,2}$ \\
$405$ & $10$ & $\gMP{3,\ldots,3}$ \\
$393$ & $8$  & $\gMP{4,4,4,4,4,4,3,3}$ \\
\multicolumn{3}{c}{$\vdots$} \\
$225$ & $2$  & $\gMP{15,15}$ \\
\multicolumn{3}{c}{$\vdots$} \\
$56$  & $2$  & $\gMP{28,2}$ \\
$29$  & $2$  & $\gMP{29,1}$ \quad (star) \\
$0$   & $1$  & $\gMP{30}$ \quad (empty network) \\
\bottomrule
\end{tabular}
\end{table}

The computation reveals a clear pattern. Along the sequence of
networks associated with the envelope vertices, those with
three or more cells are balanced or nearly balanced, with cell
sizes differing by at most one. Once the two-cell regime is reached,
the pattern changes\rev{. Starting} from $(15,15)$, the two cells become
progressively more unequal as the link cost increases, eventually
reaching $(29,1)$, which corresponds to the star. At still higher
link costs, the empty network becomes optimal. Thus, in this example,
increasing the link cost first reduces the number of cells and then,
within the two-cell regime, increases the asymmetry between them.

\rev{These observations are numerical rather than general theoretical results,
and they depend on $n$. In computations for larger $n$ at the same $p$ and
$\delta$, the number of cells again falls as the link cost increases, but
unbalanced networks appear before the two-cell regime. At $n=100$, for example,
the three-cell vertices run from $\gMP{34,33,33}$ through $\gMP{40,40,20}$ to
$\gMP{49,49,2}$, with one cell shrinking while the other two grow.}

\subsection{\texorpdfstring{\rev{The Cost of Misjudging Failure Risk}}{The Cost of Misjudging Failure Risk}}\label{sec:misspec}

\rev{The analysis so far assumes} that the survival probability is known when the
network is chosen. In practice, however, it may be uncertain or imperfectly
estimated. We therefore examine the cost of failure-risk
misspecification, that is, the loss arising when a network chosen under a
perceived survival probability $\hat p$ is evaluated under a different
true survival probability $p$. For fixed $n$, $\delta$, and $c$, let
$\rev{\bar g^{\mathrm{mp}}}(u)\in\Eff^{\mathrm{mp}}(u,\delta,c)$
denote a restricted welfare-optimal network under survival probability $u$.
We measure the welfare shortfall under the true survival probability $p$ as
\[
L(\hat p,p)
=
\frac{
\W(\rev{\bar g^{\mathrm{mp}}}(p);p,\delta,c)
-
\W(\rev{\bar g^{\mathrm{mp}}}(\hat p);p,\delta,c)
}{
\W(\rev{\bar g^{\mathrm{mp}}}(p);p,\delta,c)
}\times 100.
\]
\rev{Here $\W(g;p,\delta,c)$ denotes the welfare of $g$ evaluated at
survival probability $p$, decay factor $\delta$, and link cost $c$.}
The percentage loss is defined when $W^{*,\mathrm{mp}}(c)>0$
at the true survival probability $p$.
Thus, $L(\hat p,p)$ measures the percentage welfare loss relative to the
network that would have been chosen under the correct survival probability.
Table \ref{tab:misspec} reports this loss for $n=30$, $\delta=1/2$, and
$c=0.08$. Rows correspond to the planner's belief $\hat p$, columns to the
true survival probability $p$, and the last column reports the network
chosen under each belief. Along the diagonal, $\hat p=p$, so the welfare
loss is zero.

\begin{table}[htbp]
\centering
\caption{Welfare cost of misspecifying the survival probability at
$n=30$, $\delta=1/2$, and $c=0.08$. Entries report the percentage welfare
shortfall relative to the optimal welfare under the true $p$. Values above
$100\%$ indicate that the chosen network yields negative welfare.}
\label{tab:misspec}
\begin{tabular}{@{}lrrrrl@{}}
\toprule
belief $\hat p$ & $p=0.3$ & $p=0.5$ & $p=0.7$ & $p=0.9$
& network built \\
\midrule
$0.3$ & $0.0\%$ & $5.7\%$ & $22.6\%$ & $30.7\%$
& $\gMP{27,3}$, 81 links \\
$0.5$ & $39.5\%$ & $0.0\%$ & $18.5\%$ & $26.9\%$
& $\gMP{25,5}$, 125 links \\
$0.7$ & $728.8\%$ & $19.8\%$ & $0.0\%$ & $0.0\%$
& complete, 435 links \\
$0.9$ & $728.8\%$ & $19.8\%$ & $0.0\%$ & $0.0\%$
& complete, 435 links \\
\bottomrule
\end{tabular}
\end{table}

The table reveals an asymmetry between underestimating and overestimating
the survival probability. When the planner underestimates $p$, the chosen
network is relatively sparse, leading to forgone benefits when survival is
higher than expected. For example, the network chosen at $\hat p=0.3$,
$\gMP{27,3}$ with $81$ links, entails a $30.7\%$ welfare shortfall when
$p=0.9$. By contrast, overestimating $p$ can lead to excessive link formation.
At $\hat p=0.9$, the planner chooses the complete network with $435$ links\rev{. If} $p=0.3$, this network yields negative welfare. The additional links
continue to incur their costs even when the survival probability is too low
for their expected benefits to justify them. More generally, the entries
below the diagonal, where $\hat p>p$, tend to be larger than their
counterparts above the diagonal.

\conf{The asymmetry is not specific to this example. Underestimating survival can cost at
most the whole attainable welfare, and it never leaves the planner worse off than building
nothing.}

\begin{proposition}\label{prop:underest}
\begin{confblock}
Fix $n$, $\delta$, and $c$, and let $0<\hat p\le p<1$. Then
\begin{equation}\label{eq:underest}
\W(\bar g^{\mathrm{mp}}(\hat p);p,\delta,c)
\;\ge\;
\W(\bar g^{\mathrm{mp}}(\hat p);\hat p,\delta,c)
\;\ge\;0.
\end{equation}
Consequently, whenever $L(\hat p,p)$ is defined, that is, whenever the optimal welfare
at $p$ is positive, $0\le L(\hat p,p)\le 100$. Moreover, if $\hat p<p$, then
$L(\hat p,p)=100$ if and only if $\bar g^{\mathrm{mp}}(\hat p)$ is the empty network.
\end{confblock}
\end{proposition}

\begin{confblock}
Proposition \ref{prop:underest} also holds, with the same proof, when the planner
optimizes over all networks, since the proof uses only that the empty network is
admissible. The bound of $100\%$ is attained. By Theorem \ref{thm:extreme}, a planner
whose belief $\hat p$ is so low that $c$ exceeds the threshold \eqref{eq:chigh}
evaluated at $\hat p$ builds nothing, and thus forgoes the whole attainable welfare.
In the setting of Table \ref{tab:misspec}, this is the case at the belief $\hat p=0.2$,
for which the threshold equals $0.048<c$.
\end{confblock}

\paragraph{\rev{Managerial implications.}}
\begin{revblock}
The asymmetry has a simple source. Link costs are sunk and do not
depend on node survival, whereas the expected benefit of any fixed
network is nondecreasing in the survival probability, since the
benefit generated by a surviving pair weakly increases when
additional nodes survive. \conf{By Proposition \ref{prop:underest}, a network
chosen under a belief $\hat p\le p$ therefore yields at least the welfare the
planner anticipated, which is nonnegative because the empty network is
always available. A planner who underestimates survival may forgo part or
even all of the attainable welfare, but is never worse off than by building
nothing.} No comparable bound holds when
survival is overestimated. The planner then pays in full for links whose
benefits materialize less often than expected, and realized welfare
can fall below that of building nothing, as in the first column of
Table \ref{tab:misspec}.

For a decision maker who must commit to links before failure risk is
resolved, such as a firm contracting with suppliers, a bank extending
credit lines, or an operator installing infrastructure links, two
practical lessons follow. \conf{First, only an optimistic reliability
estimate can leave the decision maker worse off than building nothing. A
pessimistic estimate costs at most the welfare that a correctly informed
design would have attained. When the survival probability is uncertain, the
design should therefore lean toward conservative estimates, though not to
the point at which nothing is built, since the whole attainable welfare is
then forgone.} Second, a candidate
network should be evaluated not only at the estimated survival
probability but also at pessimistic values. Within the complete
multipartite class, this check is immediate from \eqref{eq:cmpB}.
In Table \ref{tab:misspec}, for example, the two-cell networks chosen
under the beliefs $\hat p=0.3$ and $\hat p=0.5$ retain positive
welfare at every true survival probability considered, whereas the
complete network, which is optimal at $p=0.7$ and $p=0.9$, falls
short of the optimal welfare by more than seven times its value
when $p=0.3$.
\end{revblock}

\subsection{Comparison \texorpdfstring{\rev{with}}{with} Heuristics}\label{sec:heuristics}

\begin{confblock}
A natural benchmark for the exact approach is provided by simple heuristic
procedures. We consider three of them, namely greedy link addition, greedy link
deletion, and multi-start local search, and we run all three on the same
instances. The experiments are conducted at $p=1/10$ and $\delta=1/2$, for
$n=6,\ldots,9$ and the eighteen cost levels
$c_j=\frac{j}{20}c_{\mathrm{high}}$, $j=2,\ldots,19$, which gives $72$ instances.

The benchmark is the unrestricted optimum $W^{*}(c)$, obtained from the
unrestricted benefit frontier by exhaustive search. For $n=6$ and $n=7$ we
enumerate all $2^{\binom{n}{2}}$ labeled graphs. For $n=8$ and $n=9$ it suffices
to enumerate the labeled graphs whose degrees are non-increasing in the node
labels, since every graph can be relabeled in this way. There are $602{,}544$
such graphs for $n=8$ and $37{,}693{,}734$ for $n=9$.
In each of the $72$ instances the optimum is attained by a complete multipartite
network. In fact, at these parameter values every vertex of the upper concave
envelope of the unrestricted benefit frontier is attained by a complete
multipartite network, so that $W^{*}(c)=W^{*,\mathrm{mp}}(c)$ for every $c>0$ and
every $n=6,\ldots,9$.
All welfare comparisons are performed using exact rational arithmetic.
\end{confblock}

\begin{revblock}
The three procedures share the same design.
Greedy link addition starts from the empty network and greedy link
deletion from the complete network. At each iteration they add,
respectively delete, the link that yields the largest increase in welfare.
Multi-start local search starts from a random network, obtained by
including each of the $\binom{n}{2}$ possible links independently with
probability $1/2$, and at each iteration makes the single-link addition
or deletion that yields the largest increase in welfare.
In all three procedures, ties are resolved according to the lexicographic
ordering of links, and the procedure terminates when no admissible move
produces a strictly positive welfare gain.
Local search is run from $60$ starting networks, generated by a Mersenne
Twister initialized with seed $0$ at each cost level, so that the same
starting networks are used throughout.
\conf{Table \ref{tab:heuristics} reports, for each $n$, how often each procedure
misses the optimum, together with the largest welfare shortfall.}
\end{revblock}

\begin{table}[htbp]
\centering
\caption{Performance of the three heuristics at $p=1/10$ and $\delta=1/2$ on
$72$ instances ($n=6,\ldots,9$ and $c_j=\frac{j}{20}c_{\mathrm{high}}$,
$j=2,\ldots,19$). Entries count the instances, or the runs, in which the optimum
is missed. The largest welfare shortfall, as a percentage of the optimal
welfare, is in parentheses.}
\label{tab:heuristics}
\small
\begin{tabular}{@{}cccccc@{}}
\toprule
$n$
& \shortstack{Complete network\\not optimal}
& \shortstack{Greedy\\addition}
& \shortstack{Greedy\\deletion}
& \shortstack{Local search,\\single runs}
& \shortstack{Local search,\\best of $60$ runs} \\
\midrule
$6$ & $4$ of $18$ & $4$ ($100\%$) & $3$ ($9.0\%$) & $144$ of $1{,}080$ ($39.1\%$) & $0$ \\
$7$ & $5$ of $18$ & $5$ ($100\%$) & $1$ ($0.3\%$) & $128$ of $1{,}080$ ($33.0\%$) & $0$ \\
$8$ & $6$ of $18$ & $6$ ($100\%$) & $1$ ($2.6\%$) & $215$ of $1{,}080$ ($67.8\%$) & $0$ \\
$9$ & $7$ of $18$ & $7$ ($100\%$) & $3$ ($3.1\%$) & $161$ of $1{,}080$ ($80.7\%$) & $0$ \\
\addlinespace
All & $22$ of $72$ & $22$ ($100\%$) & $8$ ($9.0\%$) & $648$ of $4{,}320$ ($80.7\%$) & $0$ \\
\bottomrule
\end{tabular}

\medskip
\begin{minipage}{0.95\textwidth}
\footnotesize
\emph{Notes:}
The second column counts the instances in which the complete network is not optimal.
For greedy addition, greedy deletion, and the best of $60$ local-search runs,
entries count instances, out of $18$ for each $n$.
For single local-search runs, entries count runs, out of $1{,}080$ for each $n$.
The benchmark is the unrestricted optimum $W^{*}(c)$ in every instance.
\end{minipage}
\end{table}

\begin{confblock}
Four observations follow. First, in the $50$ instances in which the complete
network is optimal, every procedure finds it. This is to be expected, since
for $c<c_{\mathrm{low}}$ adding any missing link strictly raises welfare (see the
proof of Theorem \ref{thm:extreme}). The comparison is therefore informative in
the remaining $22$ instances.

Second, greedy addition misses the optimum in all $22$ of them. When it forms
links, it stops at a network of the form $\gMP{m,1,\ldots,1}$, in which $n-m$
nodes are linked to all others and the remaining $m$ nodes are not linked to one
another. For $c>p^{2}\delta$ it forms no link at all. The first link alone changes
welfare by $2p^{2}\delta-2c<0$, and the procedure accepts only strictly improving
moves, although networks with strictly positive welfare exist. This occurs in
$7$ instances, in which the whole optimal welfare is lost.

Third, greedy deletion performs better. It misses the optimum in $8$ instances,
with shortfalls between $0.3\%$ and $9.0\%$ and a median of $2.8\%$. In each of
them it stops at a complete multipartite network that does not maximize welfare
within that class.

Fourth, the outcome of a single local-search run depends strongly on its
starting network. Of the $1{,}320$ runs on the $22$ informative instances, $648$
stop at a suboptimal network. The median shortfall of these runs is $2.9\%$ and
the largest is $80.7\%$. All but $15$ of them end at a complete multipartite
network, so convergence to such a network does not guarantee welfare optimality.
The best of the $60$ runs, by contrast, is optimal in all $72$ instances.
\end{confblock}

\begin{revblock}
None of these procedures provides a certificate of optimality.
\conf{Repeated local search found the optimum in every instance, but this could
be established only by comparison with the exhaustive benchmark, which quickly
becomes infeasible as $n$ grows.} By contrast,
the dynamic programming approach computes the exact benefit frontier within
the class of complete multipartite networks, thereby identifying at least
one welfare-optimal network in that class for every link cost.
\end{revblock}

\section{Pairwise Stability for the Restricted Class of Networks}
\label{sec:stability}
We complement the welfare analysis with the
notion of pairwise stability, one of the standard
solution concepts in the economics of network
formation introduced by \citet{JW1996}.
Pairwise stability requires that no agent wishes to sever an
existing link unilaterally and that no pair of agents has a mutually profitable
deviation to form a missing link. We use this notion to characterize the
networks that can be sustained by decentralized link formation and to compare
them with the welfare-optimal networks.
\rev{Section \ref{sec:pschar} characterizes the pairwise-stable complete multipartite
networks, Section \ref{sec:psopt} compares stability with welfare optimality, and
Section \ref{sec:existence} establishes existence.}

\subsection{\texorpdfstring{\rev{Characterization}}{Characterization}}\label{sec:pschar}
\begin{definition}[Pairwise stability, \citealp{JW1996}]
\label{def:ps}
Let $g\in\G(\N)$. The network $g$
is pairwise stable if
\begin{enumerate}[label=(\roman*),leftmargin=2.2em]
\item for every $ij\in g$\rev{, we have}
$\Pi_i(g)\ge\Pi_i(g-ij)$ and
$\Pi_j(g)\ge\Pi_j(g-ij)$\rev{,} and
\item for every $ij\notin g$\rev{,} if
$\Pi_i(g+ij)>\Pi_i(g)$, then
$\Pi_j(g+ij)<\Pi_j(g)$.
\end{enumerate}
\end{definition}

\rev{We denote by $\PS(p,\delta,c)$ the set of \rev{pairwise-stable} networks.}

\rev{We ask which complete multipartite networks are pairwise stable. Only the
candidate networks are restricted to this class. The deviations in
Definition \ref{def:ps} are unrestricted single-link additions and deletions,
even when the resulting network is no longer complete multipartite.}

For a complete multipartite network, the single-link deviations relevant
for pairwise stability have a particularly simple structure. An agent is
linked to every agent outside its own cell and to no agent within it.
Hence, an absent link can only be added between two agents in the same
cell, whereas an existing link can only be deleted between agents in
different cells. By symmetry, it therefore suffices to analyze one
representative addition for each cell and one representative deletion
for each ordered pair of distinct cells.

Two features of the model shape everything below. Adding a link never
lengthens anyone's distance to anyone, so no third party is harmed
(Proposition \ref{prop:monotone})\rev{. Moreover,} the cost $c$ falls on each endpoint,
so a link forms only if both sides gain.
We next consider the formation of a missing link within a cell.

\begin{proposition}[No addition within a cell]\label{prop:noadd}
Let $g\cong\gMP{a_1,\ldots,a_r}$ and let $i,j$ lie in the
same cell $t$, with $a_t\ge2$. Then
\begin{equation}\label{eq:addgain}
B_i(g+ij)-B_i(g)
=p^{2}\Bigl[\delta-\delta^{2}+\delta^{2}q^{n-a_t}\Bigr].
\end{equation}
Hence no pair within a cell wishes to link if and only if
$c\ge\alpha(a)$, where
\[
\alpha(a)
=p^{2}\Bigl[\delta-\delta^{2}
+\delta^{2}q^{\,n-a_{\max}}\Bigr],
\qquad
a_{\max}=\max\{a_t:a_t\ge2\},
\]
and $\alpha(a)=0$ when every cell is a singleton.
\end{proposition}

\begin{proposition}[No deletion across cells]\label{prop:nosever}
Let $g\cong\gMP{a_1,\ldots,a_r}$, let $i$ lie in cell $t$
and $j$ in cell $u\ne t$, write $\rev{x}=a_t$, $\rev{y}=a_u$ and
$w=n-\rev{x}-\rev{y}$ for the number of agents outside both cells. Then
\begin{equation}\label{eq:severloss}
\begin{split}
B_i(g)-B_i(g-ij)=\Lambda(\rev{x},\rev{y})\coloneqq\;
&\underbrace{p^{2}\delta-p^{2}\delta^{2}\bigl(1-q^{\,w}\bigr)
-p^{2}\delta^{3}q^{\,w}
\bigl(1-q^{\,\rev{x}-1}\bigr)\bigl(1-q^{\,\rev{y}-1}\bigr)}
_{\text{loss on the severed pair}}\\[6pt]
&+\underbrace{(\rev{x}-1)p^{3}\delta^{2}q^{\,n-\rev{x}-1}}
_{\text{loss of access to own cell}} .
\end{split}
\end{equation}
Hence no agent wishes to sever if and only if
$c\le\sigma(a)$, where
$\sigma(a)=\min_{t\ne u}\Lambda(a_t,a_u)$, and
$\Lambda(\rev{x},\rev{y})\ge p^{2}\delta(1-\delta)>0$
for every $\rev{x},\rev{y}$.
\end{proposition}

For $g\cong\gMP{a}$, we also write
$\alpha(g)=\alpha(a)$ and $\sigma(g)=\sigma(a)$.

\begin{remark}\label{rem:empty}
If the network consists of a single cell, say $a=(n)$, there
are no links across cells and hence no deletion constraint.
Accordingly, we define $\sigma((n))=+\infty$.
Thus, the empty network $\gE$ is pairwise stable precisely
when $c\ge p^2\delta$.
\end{remark}

\rev{Combining Propositions \ref{prop:noadd} and \ref{prop:nosever} yields the
following characterization.}

\begin{theorem}\label{prop:stabclass}
A network $g\cong\gMP{a_1,\dots,a_r}$ is pairwise stable
if and only if
\begin{equation}\label{eq:stabwindow}
\alpha(a)\le c\le\sigma(a),
\qquad
\alpha(a)=p^{2}\Bigl[
\delta-\delta^{2}+\delta^{2}q^{\,n-a_{\max}}
\Bigr],
\qquad
\sigma(a)=\min_{t\ne u}\Lambda(a_t,a_u),
\end{equation}
with $\Lambda$ given by \eqref{eq:severloss} and $a_{\max}$
the largest cell of size at least two. As above,
$\alpha(a)=0$ when all cells are singletons, and
$\sigma(a)=+\infty$ when there is only one cell.
In particular, a complete multipartite network is
pairwise stable for some cost if and only if
$\alpha(a)\le\sigma(a)$.
\end{theorem}

\begin{remark}\label{rem:binding}
\redtxt{Pairwise stability requires the cost of a link to be high enough to prevent within-cell additions, but low enough to prevent the severance of existing links. The addition constraint is set by the largest non-singleton cell and the deletion constraint by the pair of cells that minimizes $\Lambda(a_t,a_u)$.}\footnote{\citet{LJG2025} establish stability
conditions for the benchmark networks.
Our stability window for the star coincides
with theirs, while Theorem~\ref{prop:stabclass}
extends the characterization to all complete
multipartite networks.}

For example, at $n=6$, $p=1/10$, and $\delta=1/2$,
the networks $\gMP{3,1,1,1}$ and $\gMP{4,1,1}$
satisfy $\alpha(a)>\sigma(a)$ and hence are not
pairwise stable for any value of $c$. By contrast,
$\gMP{5,1}$ is pairwise stable for a specific range
of costs $\frac{19}{4000}\le c\le\frac{1}{200}$.
\end{remark}

\subsection{\texorpdfstring{\rev{Stability and Welfare Optimality}}{Stability and Welfare Optimality}}\label{sec:psopt}
\rev{Pairwise stability and welfare optimality can be incompatible.}

\begin{proposition}\label{prop:separation}
Let $n\geq3$. The interval
$\left(\max\{c_{\mathrm{mid}},p^2\delta\},
c_{\mathrm{high}}\right)$
is nonempty. For every cost $c$ in this interval,
both the set of welfare-optimal networks
$\Eff(p,\delta,c)$ and the set of \rev{pairwise-stable}
networks $\PS(p,\delta,c)$ are nonempty.
However,
\[
\Eff(p,\delta,c)\cap\PS(p,\delta,c)=\rev{\emptyset}.
\]
\end{proposition}

\redtxt{This separation holds in the full network space, not merely within the class of complete multipartite networks.}
\rev{The reason is that the center of the star bears the cost of every link but
obtains only the direct benefit $p^2\delta$ from each, whereas the social value of
a link also includes the indirect connections that it creates among peripheral agents.}

\subsection{An Existence Result}\label{sec:existence}
The preceding examples show that some complete multipartite
networks are pairwise stable over specific cost intervals,
while others are not pairwise stable for any cost.
However, these observations do not establish whether
a \rev{pairwise-stable} network exists for every link cost.
At the two extremes, existence is straightforward\rev{, since}
the complete network is pairwise stable for sufficiently
low $c$, while the empty network is pairwise stable
for $c\ge p^{2}\delta$. The key question is therefore
whether pairwise stability
can also be guaranteed for every intermediate cost.
We answer this question affirmatively using a particular
family of complete multipartite networks, which we call
quotient networks. The construction identifies
$n$ quotient networks, whose labeled realizations
form the family $\mathcal Q^{\mathrm{mp}}(\N)$.

\begin{definition}[Quotient networks]\label{def:quotient}
For each $\rev{\eta}=1,\ldots,n$, write
$n=\ell_{\rev{\eta}} \rev{\eta}+e_{\rev{\eta}}$, where
$\ell_{\rev{\eta}}=\lfloor n/\rev{\eta}\rfloor$ and $0\le e_{\rev{\eta}}<\rev{\eta}$.
We denote by $g_{[\rev{\eta}]}$ a representative of the
complete multipartite network consisting of
$\ell_{\rev{\eta}}$ cells of size $\rev{\eta}$ and, whenever $e_{\rev{\eta}}>0$,
one additional cell of size $e_{\rev{\eta}}$.
Equivalently,
\[
g_{[\rev{\eta}]}\cong
\gMP{\underbrace{\rev{\eta},\ldots,\rev{\eta}}_{\ell_{\rev{\eta}}\text{ times}},e_{\rev{\eta}}},
\]
where the residual cell is omitted if $e_{\rev{\eta}}=0$.
We define the family of quotient networks by
\[
\mathcal Q^{\mathrm{mp}}(\N)
=
\left\{
g\in\G^{\mathrm{mp}}(\N):
g\cong g_{[\rev{\eta}]}
\text{ for some }\rev{\eta}=1,\ldots,n
\right\}.
\]
In particular,
$g_{[1]}\cong\gK$,
$g_{[n-1]}\cong\gS$,
and $g_{[n]}\cong\gE$.
\end{definition}
Thus,
$\mathcal Q^{\mathrm{mp}}(\N)
\subseteq\G^{\mathrm{mp}}(\N)\subseteq\G(\N)$.

To illustrate the construction, consider $n=10$
and $\rev{\eta}=3$, yielding $g_{[3]}\cong\gMP{3,3,3,1}$,
with three cells of size $\rev{\eta}=3$ and one remaining
cell of size $1$. More generally, there is at
most one cell whose size is smaller than $\rev{\eta}$.

Two features of these networks $g_{[\rev{\eta}]}$ will be used in the proof
of the next theorem. First, by construction, the largest cell of each quotient
network $g_{[\rev{\eta}]}$ has size $\rev{\eta}$. Therefore,
Proposition \ref{prop:noadd} implies that the
threshold preventing the formation of additional
links depends only on $\rev{\eta}$ and is given by
\begin{equation}
\alpha(g_{[\rev{\eta}]})=
\begin{cases}
0, & \rev{\eta}=1,\\[4pt]
p^2\left[\delta-\delta^2+\delta^2q^{n-\rev{\eta}}\right],
& 2\le \rev{\eta}\le n.
\end{cases}
\end{equation}
In particular, $\alpha(g_{[\rev{\eta}]})$ is strictly
increasing in $\rev{\eta}$.
Second, $g_{[\rev{\eta}]}$ has at most one cell smaller
than $\rev{\eta}$, so, among every pair of distinct cells,
at least one has size $\rev{\eta}$.

By Theorem \ref{prop:stabclass}, a complete multipartite
network $g$ is pairwise stable precisely when
$c\in[\alpha(g),\sigma(g)]$. We refer to this
possibly empty interval as the \emph{stability window}
of $g$. \rev{The stability windows of consecutive quotient networks meet or overlap and
jointly cover the entire range of link costs, which guarantees the existence
of a pairwise-stable network for every $c>0$.}

\begin{theorem}\label{thm:existence}
For every $n\ge3$, $p,\delta\in(0,1)$ and $c>0$,
the family of quotient networks contains a
\rev{pairwise-stable} network. In particular,
$\mathcal Q^{\mathrm{mp}}(\N)\cap
\PS(p,\delta,c)\ne\emptyset$.
Indeed, the stability windows of the quotient
networks satisfy
\begin{equation}\label{eq:chain}
\alpha(g_{[\rev{\eta}+1]})\le\sigma(g_{[\rev{\eta}]}),
\qquad \rev{\eta}=1,\ldots,n-1,
\end{equation}
so that
$\bigcup_{\rev{\eta}=1}^{n}
[\alpha(g_{[\rev{\eta}]}),\sigma(g_{[\rev{\eta}]})]
=[0,\infty)$.
\end{theorem}

\rev{\conf{The key observation behind \eqref{eq:chain} is that two agents who sever their link in $g_{[\rev{\eta}]}$ are left with at most $n-\rev{\eta}-1$ common neighbors, no more than two unlinked agents in the largest cell of $g_{[\rev{\eta}+1]}$, so a direct link is worth at least as much to the former as to the latter. The proof verifies that this advantage is not offset by the longer routes that remain after severance.}}

Since pairwise stability of quotient networks is
evaluated with respect to all feasible single-link
deviations in the unrestricted network space
$\G(\N)$, Theorem \ref{thm:existence}
immediately yields the following general
existence result.

\begin{corollary}\label{cor:existence}
$\PS(p,\delta,c)\ne\emptyset$ for every
$n\ge3$, $p,\delta\in(0,1)$ and $c>0$.
\end{corollary}

The family of quotient networks provides a sufficient
construction for the existence of \rev{pairwise-stable}
networks, but does not exhaust all \rev{pairwise-stable}
networks. For example, when $n=10$ and
$p=\delta=0.9$, the network $\gMP{4,3,3}$
is pairwise stable over a non-degenerate interval
of costs, although it does not belong to
$\mathcal Q^{\mathrm{mp}}(\N)$.

\section{Concluding Remarks}\label{sec:extensions}
In this paper, we study optimal network design under independent node failures, characterizing welfare-optimal networks and developing a polynomial-time algorithm for welfare maximization within the class of complete multipartite networks. We also investigate decentralized network formation, characterizing pairwise-stable networks when nodes act as strategic economic agents.
Three extensions offer promising directions for future research.

\medskip\noindent
\emph{Heterogeneous failure probabilities.}
Allowing nodes to have different survival probabilities could
substantially change optimal networks. More reliable
nodes may become natural hubs, potentially favoring asymmetric
\rev{core-periphery} networks.
Heterogeneity would also challenge our computational approach,
which relies on the exchangeability of nodes within each cell.
Developing an alternative representation that preserves
computational tractability is therefore an important open problem.

\medskip\noindent
\emph{Correlated failures.}
Common shocks may affect several nodes simultaneously, changing
the value of alternative paths and network redundancy.
Exchangeable survival distributions offer a potentially tractable
extension, since the expected-benefit formula for complete
multipartite networks can be adapted by replacing the relevant
independent-survival probabilities with their correlated
counterparts. More general dependence structures may require
different analytical and computational methods.

\medskip\noindent
\emph{\rev{Cost transfers.}}
Allowing agents to share link costs could support connections
that are not individually profitable under equal cost sharing,
potentially reducing the gap between \rev{welfare optimality} and pairwise
stability. However, transfers alone cannot overcome the
limitations of single-link deviations when moving between
optimal networks requires coordinated changes to several
links. Addressing this issue would require extending the
stability analysis to coordinated deviations.

\section*{Acknowledgments}
Francesco Ciardiello acknowledges partial financial
support from the University of Salerno's
\textit{Fondo di Ateneo per la Ricerca di Base (FARB)}.
This support was provided through projects
300399FRB23CIARD and 300399FRB24CIARD,
both entitled ``Game Theory and Applications.''

\section*{Declaration of Competing Interests}
The authors declare that they have no known competing financial interests or personal relationships that
could have appeared to influence the work reported in this paper.


\appendix
\section{\texorpdfstring{\rev{Proofs}}{Proofs}}\label{app:proofs}
\rev{This appendix collects the proofs of all results, in the order in which they are stated in the text.}

\subsection*{\rev{Proofs for Section \ref{sec:classes}}}

\begin{proof}[\rev{Proof of Lemma \ref{lem:cmpform}}]
The link count follows by counting pairs of nodes
in different cells. For the expected benefit, let
$X_t$ denote the number of surviving nodes in cell $t$.
Random variables $X_t$ are independent, with
$X_t\sim\mathrm{Bin}(a_t,p)$.

Surviving nodes in different cells are directly connected.
Their expected contribution is therefore
\[
\delta\,\mathbb{E}\left[\sum_{t\ne u}X_tX_u\right]
=
2p^2\delta\,|g|.
\]

Two surviving nodes in the same cell $t$ are connected
at distance two if and only if at least one node outside
that cell survives. This event has probability
$1-q^{n-a_t}$ and is independent of $X_t$.
Since $\mathbb{E}[X_t(X_t-1)]=a_t(a_t-1)p^2$,
their expected contribution is
\[
p^2\delta^2
\sum_t a_t(a_t-1)\left(1-q^{n-a_t}\right).
\]
Adding the two contributions yields \eqref{eq:cmpB}.
Finally, a node $i$ in cell $t$ has $n-a_t$ direct
neighbors and $a_t-1$ nodes in its own cell, each
reachable at distance two whenever at least one node
outside the cell survives. This gives the individual
benefit in \eqref{eq:cmpayoff}. Subtracting the cost
of the $n-a_t$ incident links yields the corresponding
payoff.
\end{proof}

\begin{proof}[\rev{Proof of Proposition \ref{prop:monotone}}]
Adding a link cannot increase the distance between
any two surviving nodes, regardless of which other
nodes fail. Since benefits are decreasing in distance,
$B_m(g+ij)\ge B_m(g)$ for every $m\in\N$.
Moreover, whenever both $i$ and $j$ survive, their
distance decreases from at least two (or infinity)
to one. Since this event has probability $p^2>0$
and $\delta\in(0,1)$, both nodes obtain a strictly
positive increase in expected benefits.
Hence $B_i(g+ij)>B_i(g)$ and
$B_j(g+ij)>B_j(g)$, implying
$\Ben(g+ij)>\Ben(g)$.
\end{proof}

\begin{proof}[\rev{Proof of Proposition \ref{prop:hull}}]
Write $\rev{\tau}=2c>0$. By \eqref{eq:welfare},
welfare maximization over $\mathcal A$ is equivalent
to maximizing $\Ben_{\mathcal A}^*(k)-\rev{\tau}k$
over $k\in\mathcal K_{\mathcal A}$.

Proposition \ref{prop:monotone} implies that the
complete network $\gK$ yields strictly greater
aggregate expected benefits than any other network.
Since $\gK$ belongs to both admissible classes
and has the largest feasible link count, the final
point of either benefit frontier lies strictly
above all preceding points. Consequently, the
last segment of its upper concave envelope has
strictly positive slope, and concavity implies
that all preceding segments also have strictly
positive slopes.

A link count $k$ is welfare-optimal at cost $c$
if and only if $(k,\Ben_{\mathcal A}^*(k))$
maximizes the linear functional
$(x,y)\mapsto y-\rev{\tau}x$ over the benefit frontier.
Since all segments of its upper concave envelope
have strictly positive slopes, this occurs
for some $\rev{\tau}>0$ if and only if the corresponding
frontier point lies on that envelope.
\end{proof}

\subsection*{\rev{Proofs for Section \ref{sec:efficiency}}}

\begin{proof}[\rev{Proof of Theorem \ref{thm:perlink}}]
For each survival state $S\subseteq\N$, define
\[
N_{\delta}(g|_S)
=
\sum_{\substack{i,j\in S\\i\ne j}}
\delta^{\,\dist{g|_S}{i}{j}},
\qquad
\delta^{\infty}=0.
\]
This quantity measures the aggregate connection benefit
generated by the residual network in state $S$,
before taking expectations over node survival.
By \eqref{eq:payoff},
\[
\Ben(g)
=
\sum_{S\subseteq\N}
\lambda(S)N_{\delta}(g|_S).
\]

We first establish an auxiliary bound for each
residual network. Let $S\subseteq\N$ with
$|S|=\rev{\kappa}\ge2$, and let
\[
m_S
=
\bigl|\{ij\in g:i,j\in S\}\bigr|
\]
be the number of links in $g|_S$.
We claim that
\begin{equation}\label{eq:counting}
N_{\delta}(g|_S)
\le
m_S\bigl[2\delta+(\rev{\kappa}-2)\delta^2\bigr],
\end{equation}
with equality, when $m_S>0$, if and only if
$g|_S$ is a star spanning $S$.

To prove this bound, observe that the $2m_S$
ordered pairs joined by a link contribute
$2m_S\delta$. Let $C_2$ denote the number of
ordered pairs of distinct non-adjacent nodes
belonging to the same connected component
of $g|_S$.
Since every such pair is at distance at least two,
\begin{equation}\label{eq:c2bound}
N_{\delta}(g|_S)
\le
2m_S\delta+C_2\delta^2.
\end{equation}

Let the nontrivial connected components of $g|_S$
have $\rev{b_1,\ldots,b_{\omega}}$ vertices and
$m_1,\ldots,m_{\rev{\omega}}$ links, respectively.
Then
\[
C_2
=
\sum_{t=1}^{\rev{\omega}}
\bigl[\rev{b_t}(\rev{b_t}-1)-2m_t\bigr].
\]
Since each connected component satisfies
$m_t\ge \rev{b_t}-1$, we have
\[
\rev{b_t}(\rev{b_t}-1)-2m_t
\le
m_t(\rev{b_t}-2)
\le
m_t(\rev{\kappa}-2).
\]
Consequently,
\[
C_2
\le
\sum_{t=1}^{\rev{\omega}}m_t(\rev{\kappa}-2)
=
m_S(\rev{\kappa}-2).
\]
Substituting into \eqref{eq:c2bound} gives
\eqref{eq:counting}.

When $m_S>0$, equality in this bound requires
a single connected component spanning $S$,
with exactly $\rev{\kappa}-1$ links.
Moreover, every non-adjacent pair must be
at distance two.
Thus $g|_S$ is a spanning tree of diameter
at most two, and hence a star.
Conversely, a spanning star attains equality
in \eqref{eq:counting}.

We now apply the bound to the expected
aggregate benefit. Since states with
$|S|\le1$ contribute nothing,
\[
\begin{aligned}
\Ben(g)
&=
\sum_{\substack{S\subseteq\N\\|S|\ge2}}
\lambda(S)N_{\delta}(g|_S)\\
&\le
\sum_{\substack{S\subseteq\N\\|S|\ge2}}
\lambda(S)m_S
\bigl[2\delta+(|S|-2)\delta^2\bigr].
\end{aligned}
\]

Using
\[
m_S
=
\sum_{ij\in g}\ind{i,j\in S},
\]
and interchanging the finite sums, we obtain
\[
\Ben(g)
\le
\sum_{ij\in g}
\sum_{S\supseteq\{i,j\}}
\lambda(S)
\bigl[2\delta+(|S|-2)\delta^2\bigr].
\]

For each fixed link $ij$, conditional on
both endpoints surviving, the number of
additional surviving nodes follows a
$\mathrm{Bin}(n-2,p)$ distribution.
Therefore,
\[
\begin{aligned}
&\sum_{S\supseteq\{i,j\}}
\lambda(S)
\bigl[2\delta+(|S|-2)\delta^2\bigr]\\
&\qquad=
p^2\left[2\delta+(n-2)p\delta^2\right]\\
&\qquad=
2p^2\delta+(n-2)p^3\delta^2.
\end{aligned}
\]

Since this expression is independent of
the chosen link, it follows that
\[
\Ben(g)
\le
|g|\bigl[2p^2\delta+(n-2)p^3\delta^2\bigr].
\]

Dividing by $|g|>0$ establishes the
inequality in \eqref{eq:perlink}.

It remains to characterize equality.
If $g=\gS$,
then in every survival state the residual
network is either a star spanning the
surviving nodes or an empty network.
Hence equality holds in \eqref{eq:counting}
for every state, and the upper bound
in \eqref{eq:perlink} is attained.

Conversely, suppose equality holds in
\eqref{eq:perlink}. Since
$\lambda(S)>0$ for every $S\subseteq\N$,
equality must hold in \eqref{eq:counting}
for every state with $|S|\ge2$.
In particular, taking $S=\N$ and using
$|g|>0$, the equality condition established
above implies that $g$ is a spanning star.
Thus $g=\gS$.
\end{proof}

\begin{proof}[\rev{Proof of Theorem \ref{thm:extreme}}]
For part (i), consider any incomplete network $g$ and a
missing link $ij$. Adding $ij$ increases aggregate expected
benefits by at least
$
2p^2\left[\delta-\delta^2+\rev{q^{\,n-2}}\delta^2\right]
=
2c_{\mathrm{low}}.
$
To justify this bound, conditional on both endpoints surviving,
their connection before the addition contributes at most $\delta^2$
if at least one other node survives, and zero otherwise.
Since the probability that no other node survives is
$\rev{q^{\,n-2}}$, the aggregate expected gain on the two ordered
pairs $(i,j)$ and $(j,i)$ is at least
$2p^2[\delta-\delta^2(1-\rev{q^{\,n-2}})]=2c_{\mathrm{low}}$.
All other ordered-pair contributions weakly increase by
Proposition \ref{prop:monotone}.

Hence, if $c<c_{\mathrm{low}}$, the benefit increase exceeds
the aggregate link cost $2c$, so every incomplete network
can be strictly improved by adding a link. Thus $\gK$
is the unique welfare optimum. At $c=c_{\mathrm{low}}$,
adding any missing link weakly increases welfare,
so $\gK$ remains welfare-optimal.

Now consider a network $\gKm$ obtained by removing
a single link from $\gK$. The resulting benefit loss is
exactly
$\Ben(\gK)-\Ben(\gKm)=
2p^2\left[\delta-\delta^2+\rev{q^{\,n-2}}\delta^2\right]
=2c_{\mathrm{low}}.$
Consequently,
\[
\W(\gK)-\W(\gKm)
=
2(c_{\mathrm{low}}-c).
\]
Thus, for $c>c_{\mathrm{low}}$,
$\gKm$ strictly outperforms $\gK$,
so the complete network cannot be welfare-optimal.
\rev{Finally}, at $c=c_{\mathrm{low}}$, the two networks are
co-optimal.

For part (ii), Theorem \ref{thm:perlink} gives
$
\W(g)
\le
|g|\left[2p^2\delta+(n-2)p^3\delta^2-2c\right]
=
2|g|(c_{\mathrm{high}}-c).
$
Thus every nonempty network has strictly negative welfare when
$c>c_{\mathrm{high}}$, whereas $\W(\gE)=0$. At
$c=c_{\mathrm{high}}$, equality among nonempty networks holds, by
Theorem \ref{thm:perlink}, if and only if $g$ is a star. Hence
$
\Eff(p,\delta,c_{\mathrm{high}})=
\rev{\{\gE,\gS\}}.
$
Finally, for $c<c_{\mathrm{high}}$, we have
$
\W(\gS)
=2(n-1)(c_{\mathrm{high}}-c)>0
=\W(\gE).
$
Hence, the empty network cannot be welfare-optimal
for $c<c_{\mathrm{high}}$.
\end{proof}

\begin{proof}[\rev{Proof of Theorem \ref{thm:star}}]
For any $k>n-1$, the star yields at least as much welfare as every $k$-link
network if and only if
\[
c\ge
\frac{\Ben^{*}(k)-\Ben(\gS)}
     {2(k-(n-1))}.
\]
Hence, by the definition of $c_{\mathrm{mid}}$, the star strictly outperforms
every network with more than $n-1$ links when $c>c_{\mathrm{mid}}$, while at
$c=c_{\mathrm{mid}}$ it ties with at least one such network.

Now consider $0<k<n-1$. By Theorem \ref{thm:perlink},
\[
\frac{\Ben^{*}(k)}{k}
<
\frac{\Ben(\gS)}{n-1}
=
2c_{\mathrm{high}},
\]
and therefore
\[
\begin{aligned}
\W(\gS)-\bigl[\Ben^{*}(k)-2ck\bigr]
&>
(n-1-k)
\left[
\frac{\Ben(\gS)}{n-1}-2c
\right] \\
&=
2(n-1-k)(c_{\mathrm{high}}-c)>0
\end{aligned}
\]
whenever $c<c_{\mathrm{high}}$. The empty network is also strictly worse,
since
$\W(\gS)=2(n-1)(c_{\mathrm{high}}-c)>0=\W(\gE).$
Thus, below $c_{\mathrm{high}}$, every network with fewer than $n-1$ links
yields strictly lower welfare than the star.
Finally, for every $k>n-1$, Theorem \ref{thm:perlink} implies
\[
\frac{\Ben^{*}(k)-\Ben(\gS)}
     {2(k-(n-1))}
<
\frac{\Ben(\gS)}{2(n-1)}
=
c_{\mathrm{high}}.
\]
Taking the maximum over $k>n-1$ gives
$c_{\mathrm{mid}}<c_{\mathrm{high}}.$
It follows that, for $c_{\mathrm{mid}}<c<c_{\mathrm{high}}$, every network with
a number of links different from $n-1$ yields strictly lower welfare than
the star. Among networks with $n-1$ links, Theorem \ref{thm:perlink} implies
that only the spanning stars attain the maximal benefit per link. Hence
$
\Eff(p,\delta,c)=\rev{\{\gS\}}.$
At $c=c_{\mathrm{mid}}$, the maximum defining $c_{\mathrm{mid}}$ is attained, so
the star ties with at least one benefit-maximizing network having more than
$n-1$ links. Since $c_{\mathrm{mid}}<c_{\mathrm{high}}$, all networks with fewer
than $n-1$ links yield strictly lower welfare.
Thus, at \(c=c_{\mathrm{mid}}\), the star network is welfare-optimal but is not the only welfare-optimal network.
\end{proof}

\begin{proof}[\rev{Proof of Proposition \ref{prop:thresholds}}]
The inequality $c_{\mathrm{mid}}<c_{\mathrm{high}}$ follows from
Theorem \ref{thm:star}. It remains to compare
$c_{\mathrm{low}}$ and $c_{\mathrm{mid}}$.
Consider the network
obtained from a spanning star by adding a link between two of its leaves.
This network has $n$ links. The additional link increases expected benefit
by $2p^2(\delta-p\delta^2)$\rev{, since,} conditional on the two leaves surviving, their
connection changes from a distance-two connection requiring survival of
the center to a direct connection. Hence, by \rev{\eqref{eq:cmid}},
$c_{\mathrm{mid}}\geq p^2(\delta-p\delta^2)$.
Using \rev{\eqref{eq:clow}}, we have
\[
\begin{aligned}
p^2(\delta-p\delta^2)-c_{\mathrm{low}}
&=
p^2\delta^2
\left[\rev{q}-\rev{q^{\,n-2}}\right].
\end{aligned}
\]
For $0<p<1$, this expression is nonnegative for $n\geq3$ and strictly
positive for $n>3$. Thus
$c_{\mathrm{low}}\leq c_{\mathrm{mid}}$, with strict inequality whenever
$n>3$.
If $n=3$, the only network with more than $n-1=2$ links is the complete
network. Hence the network considered above is the only network entering
the maximum that defines $c_{\mathrm{mid}}$, and the preceding inequality holds
with equality. Therefore $c_{\mathrm{low}}=c_{\mathrm{mid}}$ when $n=3$.
\end{proof}

\begin{proof}[\rev{Proof of Corollary \ref{cor:linkcounts}}]
For $c<c_{\mathrm{high}}$, the proof of
Theorem \ref{thm:star} shows that every network
with $0\leq |g|\leq n-2$ yields strictly lower
welfare than a star.
For $c\geq c_{\mathrm{high}}$,
Theorem \ref{thm:extreme} establishes that the only
welfare-optimal networks are the empty network and,
at $c=c_{\mathrm{high}}$, the stars,
which have $0$ and $n-1$ links, respectively.
Therefore, no network with
$1\leq |g|\leq n-2$ can be welfare-optimal
for any $c$.
\end{proof}

\begin{proof}[\rev{Proof of Corollary \ref{cor:intermediate}}]
Since $n\geq 4$, we have
$c_{\mathrm{low}}<c_{\mathrm{mid}}<c_{\mathrm{high}}$.
For $c\in(c_{\mathrm{low}},c_{\mathrm{mid}})$,
the complete network is not welfare-optimal,
since $c>c_{\mathrm{low}}$.
Moreover, by the definition of $c_{\mathrm{mid}}$,
there exists a network with more than $n-1$ links
yielding strictly higher welfare than the star.
Since $c<c_{\mathrm{high}}$, the empty network
also yields strictly lower welfare than the star.

Corollary \ref{cor:linkcounts} excludes networks
with $1\leq |g|\leq n-2$, while
Theorem \ref{thm:perlink} implies that every
network with exactly $n-1$ links other than
a star yields strictly lower welfare than the star.

Hence, no network with fewer than $n$ links
can be welfare-optimal.
Since the complete network is also excluded,
every welfare-optimal network $g$ must satisfy
$|g|\in\mathcal{I}$.
\end{proof}

\begin{proof}[\rev{Proof of Theorem \ref{prop:matching}}]
Let $g=\gK\setminus M$, where $M$ is a set
of $\rev{\rho}$ links. By the lower bound on the
benefit increase from adding a missing link,
\[
\Ben(\gK)-\Ben(g)\geq 2\rev{\rho} c_{\mathrm{low}}.
\]

Suppose first that $M$ is a matching.
For each missing link $ij$, if both endpoints
survive, their distance in the residual network
is two whenever at least one other node survives.
Otherwise, they are disconnected.
Hence, the benefit loss associated with each
missing link is exactly $2c_{\mathrm{low}}$,
and therefore
\[
\Ben(\gK)-\Ben(g)=2\rev{\rho} c_{\mathrm{low}}.
\]

Conversely, suppose that $M$ is not a matching.
Then two missing links, say $ij$ and $i\rev{z}$,
share a node.
Consider the realization in which exactly
$i$, $j$, and $\rev{z}$ survive.
In this realization, $i$ is disconnected
from both $j$ and $\rev{z}$, so the benefit loss
is strictly greater than the lower bound
used above.
Since this realization has probability
$p^3\rev{q^{\,n-3}}>0$, we obtain
\[
\Ben(\gK)-\Ben(g)>2\rev{\rho} c_{\mathrm{low}}.
\]

Thus,
\[
\Ben(\gK)-\Ben(g)=2\rev{\rho} c_{\mathrm{low}}
\]
if and only if $M$ is a matching.
At $c=c_{\mathrm{low}}$, we have
\[
\W(\gK)-\W(g)
=
\Ben(\gK)-\Ben(g)-2\rev{\rho} c_{\mathrm{low}}.
\]
Therefore, equality holds if and only if
$M$ is a matching.
The welfare-optimal networks are
thus precisely those represented by \rev{graphs}
obtained by removing matchings from $\gK$.
Since all matchings of the same cardinality
yield isomorphic networks, each matching size
corresponds to one network.
As $\rev{\rho}$ ranges from $0$ to $\chi$, there are
exactly $\chi+1$ distinct welfare-optimal
networks.
\end{proof}

\begin{proof}[\rev{Proof of Theorem \ref{prop:right_low}}]
By Theorem \ref{prop:matching},
at $c=c_{\mathrm{low}}$ the welfare-optimal
networks are exactly those represented
by \rev{graphs} obtained by removing a matching.
Let $g_{\rev{\rho}}$ be a network obtained by removing
a matching of cardinality $\rev{\rho}$, where
$0\leq \rev{\rho}\leq\chi$.
For every such network,
\[
\Ben(g_{\rev{\rho}})
=
\Ben(\gK)-2\rev{\rho} c_{\mathrm{low}}.
\]
Consequently,
\[
\W(g_{\rev{\rho}})
=
\W(\gK)+2\rev{\rho}(c-c_{\mathrm{low}}).
\]
For $c>c_{\mathrm{low}}$, this expression
is strictly increasing in $\rev{\rho}$.
Hence, among all networks that are
welfare-optimal at $c=c_{\mathrm{low}}$,
those represented by \rev{graphs} with $\rev{\rho}=\chi$
yield strictly higher welfare than those
with $\rev{\rho}<\chi$.

It remains to exclude all other networks.
By Theorem \ref{prop:matching},
every network not obtained by removing
a matching has strictly lower welfare
at $c=c_{\mathrm{low}}$ than a network
obtained by removing a maximum matching.
Since the set of networks is finite and
the welfare of each network is continuous
in $c$, there exists $\varepsilon>0$
such that these strict inequalities
continue to hold for
$c\in(c_{\mathrm{low}},
c_{\mathrm{low}}+\varepsilon)$.
Therefore, throughout this interval, \rev{the
unique welfare-optimal network is the one}
obtained from $\gK$ by removing a maximum matching.
\end{proof}

\begin{proof}[\rev{Proof of Theorem \ref{prop:never_optimal}}]
Let $g=\gK\setminus M$, where $|M|=\rev{\rho}$
and $M$ is not a matching.
Since $\rev{\rho}\leq\lfloor n/2\rfloor$,
there exists a matching $M'$ of cardinality
$\rev{\rho}$ in $\gK$.
Consider the network $g'=\gK\setminus M'$.
By Theorem \ref{prop:matching},
at $c=c_{\mathrm{low}}$, $g'$ is
welfare-optimal, whereas $g$ is not.
Hence,
\[
\W(g')>\W(g)
\]
at $c=c_{\mathrm{low}}$.
Since $g$ and $g'$ have the same number
of links, their welfare difference
is independent of $c$.
Therefore,
$\W(g')>\W(g)$
for every $c>0$.
Thus, $g$ cannot be welfare-optimal
for any link cost.
\end{proof}

\begin{proof}[\rev{Proof of Proposition \ref{prop:n4}}]
By Corollary \ref{cor:linkcounts}, networks with
one or two links are never welfare-optimal.
Theorem \ref{thm:perlink} implies that, among
three-link networks, only stars can be
welfare-optimal.
Up to isomorphism, the only four-link networks
are the cycle $\gC$ and the paw network $g^p$.
Theorem \ref{prop:never_optimal} excludes
the paw network from welfare optimality.
Thus, the only possible welfare-optimal networks
are $\gK$, $\gKm$, $\gC$, $\gS$,
and $\gE$.
By Theorem \ref{prop:matching},
\[
\Ben(\gK)-\Ben(\gKm)
=
\Ben(\gKm)-\Ben(\gC)
=
2c_{\mathrm{low}}.
\]
Hence the benefit-frontier points corresponding
to $\gC$, $\gKm$, and $\gK$ are collinear,
with slope $2c_{\mathrm{low}}$.
Moreover,
\[
\begin{aligned}
\Ben(\gC)
&=8p^2\delta+4p^3(2-p)\delta^2,\\
\Ben(\gS)
&=6p^2\delta+6p^3\delta^2.
\end{aligned}
\]
The slope of the segment joining the star
and the cycle is therefore
\[
\Ben(\gC)-\Ben(\gS)
=
2\left[p^2\delta
+p^3(1-2p)\delta^2\right]
=:2\bar c.
\]
Since
\[
\bar c-c_{\mathrm{low}}
=
3p^3\rev{q}\delta^2>0,
\]
the definition \rev{\eqref{eq:cmid}} of $c_{\mathrm{mid}}$ gives
\[
c_{\mathrm{mid}}
=
\max\left\{
\bar c,\,
\frac{\bar c+c_{\mathrm{low}}}{2},\,
\frac{\bar c+2c_{\mathrm{low}}}{3}
\right\}
=
\bar c.
\]

Finally, the segment joining $\gE$ and
$\gS$ has slope $2c_{\mathrm{high}}$,
with
\[
c_{\mathrm{high}}-c_{\mathrm{mid}}
=
2p^4\delta^2>0.
\]

Thus, the upper concave envelope of the
benefit frontier consists of the segments
joining, in increasing order of link counts,
$\gE$, $\gS$, $\gC$, and $\gK$,
with $\gKm$ lying on the last segment.
Their slopes are, respectively,
$2c_{\mathrm{high}}$,
$2c_{\mathrm{mid}}$, and
$2c_{\mathrm{low}}$.
By Proposition \ref{prop:hull}, these
supported link counts are precisely those
that can be welfare-optimal for positive
link costs. Since welfare maximization
selects the points supported by a line
with slope $2c$, the consecutive slopes
identify the switching costs and establish
all the welfare-optimality regions and
co-optimality cases stated in the proposition.
\end{proof}

\subsection*{\rev{Proofs for Section \ref{sec:computation}}}

\begin{proof}[\rev{Proof of Proposition \ref{prop:n4mp}}]
By Proposition \ref{prop:n4}, the only networks
that can be welfare-optimal are $\gK$, $\gKm$,
$\gC$, $\gS$, and $\gE$.
All are complete \rev{multipartite, since}
\rev{$\gK\cong\gMP{1,1,1,1}$,
$\gKm\cong\gMP{2,1,1}$,
$\gC\cong\gMP{2,2}$,
$\gS\cong\gMP{1,3}$, and
$\gE\cong\gMP{4}$.}
Hence the two optimal correspondences coincide.
\end{proof}

\begin{proof}[\rev{Proof of Proposition \ref{prop:chord_exclusion}}]
Recall that $\W(g)=\Ben(g)-2c|g|$.
Since $m=\theta m^-+(1-\theta)m^+$,
condition \eqref{eq:chord_exclusion} implies that,
for every $c>0$,
\[
\begin{aligned}
\W(g)
&<
\theta\W(g^-)+(1-\theta)\W(g^+)\\
&\leq
\max\{\W(g^-),\W(g^+)\}.
\end{aligned}
\]
Hence at least one of $g^-$ and $g^+$ yields
strictly higher welfare than $g$.
\end{proof}

\begin{proof}[\rev{Proof of Proposition \ref{prop:n5}}]
For $n=5$, the benefit-maximizing networks
with $6$, $8$, and $9$ links are complete multipartite.
The only potentially problematic link counts are
therefore $5$ and $7$.
With five links, the benefit maximum is attained
by either the five-cycle $\gC$ or $\gA{5}$,
depending on $p$ and $\delta$, with both being
maximizers when their benefits coincide.
The network $\gA{5}$ consists of a four-cycle
with an additional pendant vertex.
With seven links, the benefit maximum is attained
by $\gA{7}$, obtained from the complete network
by removing two links sharing a vertex and a third
link disjoint from both.
These three non-complete-multipartite networks
are illustrated in Figure \ref{fig:n5frontier}.

\begin{figure}[htbp]
    \centering
    \includegraphics[width=0.95\textwidth]{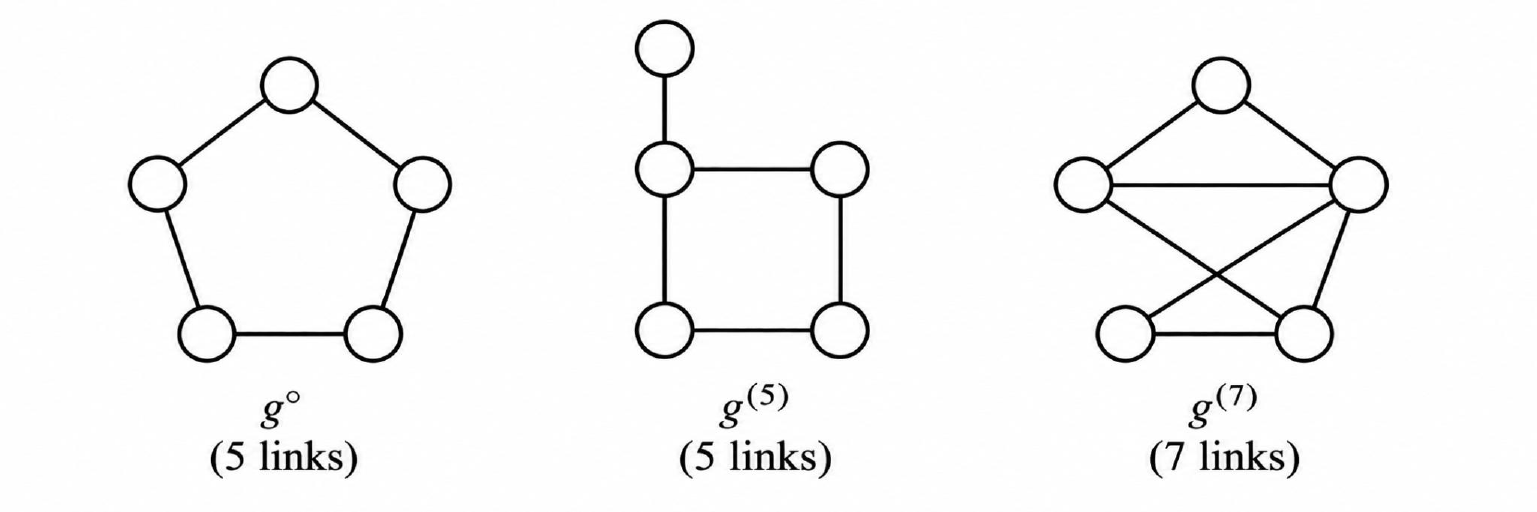}
    \caption{Non-complete-multipartite networks attaining
    the unrestricted benefit frontier for $n=5$ and some
    admissible values of $p$ and $\delta$.}
    \label{fig:n5frontier}
\end{figure}
We exclude these networks using
Proposition \ref{prop:chord_exclusion}.
For $\gA{7}$, the complete multipartite networks
$\gMP{2,3}$ and $\gMP{2,2,1}$ have six and eight
links, respectively. The corresponding chord,
with $\theta=1/2$, gives
\[
\Ben(\gA{7})
-\frac12\left[
\Ben(\gMP{2,3})+\Ben(\gMP{2,2,1})
\right]
=
-p^3\delta^2\rev{q^{2}}<0.
\]
Hence no network with seven links can be
welfare-optimal. For the five-link networks, consider the chord
joining $\gS$ and $\gMP{2,3}$, with $\theta=1/2$.
Let
\[
L_5=\frac12\left[
\Ben(\gS)+\Ben(\gMP{2,3})
\right].
\]
For $\gA{5}$,
\[
\Ben(\gA{5})-L_5
=
-p^3\delta^2
\left[
p^2-2p+3-2\delta p(2-p)
\right]<0,
\]
since
\[
p^2-2p+3-2\delta p(2-p)
=
3\rev{q^{2}}+2p(2-p)(1-\delta)>0.
\]
Thus, $\gA{5}$ cannot be welfare-optimal.

For $\gC$,
\begin{equation}
\Ben(\gC)-L_5
=
p^3\delta^2\rev{q}
\left[p(1+10\delta)-5\right].
\end{equation}
This difference is strictly negative whenever
$p(1+10\delta)<5$.
Under this condition, neither of the two
benefit-maximizing networks with five links
can be welfare-optimal. Since the same holds
unconditionally for seven links, all remaining
potentially optimal networks are complete
multipartite.
This completes the proof.
\end{proof}

\begin{proof}[\rev{Proof of Proposition \ref{prop:n6}}]
For $n=6$, the unrestricted benefit frontier
includes seven identified non-complete-multipartite
networks\rev{. These are} $\gA{6,1}$, $\gA{6,2}$,
$\gA{6,3}$, and $\gC$ with six \rev{links,}
$\gA{7,1}$ and $\gA{7,2}$ with seven \rev{links,}
and $\gA{10}$ with ten links.
These networks are illustrated in
Figure \ref{fig:n6frontier}.

\begin{figure}[htbp]
\centering
\includegraphics[width=0.95\textwidth]{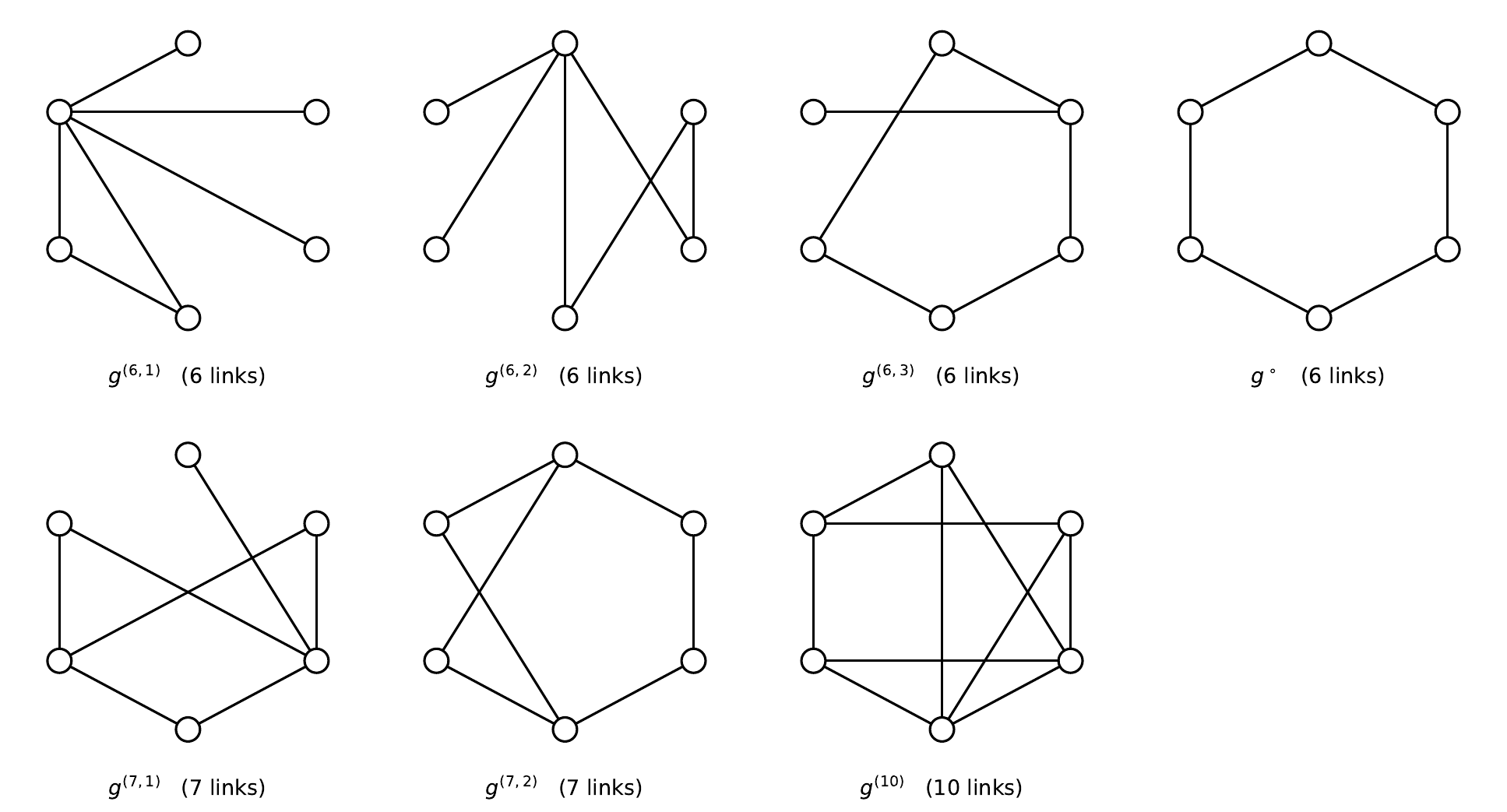}
\caption{Non-complete-multipartite networks identified
on the unrestricted benefit frontier for $n=6$
and some admissible values of $p$ and $\delta$.}
\label{fig:n6frontier}
\end{figure}

We first exclude two six-link networks
from the benefit frontier.
Observe that $Q(p,\delta)>0$ implies
\[
p\delta<
\frac{2(p^2-3p+6)}{9(3-p)}
<\frac49<\frac12.
\]
For $\gA{6,3}$, a direct computation gives
\[
\Ben(\gA{6,2})-\Ben(\gA{6,3})
=
2p^3\delta^2\rev{q(1-2p\delta)}(p\delta+2).
\]
Since \rev{$q>0$ and $1-2p\delta>0$},
the difference is strictly positive.

Similarly,
$
\Ben(\gA{6,2})-\Ben(\gC)=
2p^3\delta^2H(p,\delta),
$
where
\[
H(p,\delta)
=
\rev{-6p^2q}\delta^2
+p(3p^2-2p-2)\delta
-2p+3.
\]
For each $p\in(0,1)$, $H(p,\delta)$ is concave
in $\delta$. Moreover,
\[
H(p,0)=3-2p>0
\]
and
\[
H\left(p,\frac4{9p}\right)
=
\frac{36p^2-46p+25}{27}>0.
\]
By concavity, $H(p,\delta)>0$ for
$0<\delta<4/(9p)$.
Thus, under $Q(p,\delta)>0$, neither
$\gA{6,3}$ nor $\gC$ can be
welfare-optimal.

We exclude the remaining networks using
Proposition \ref{prop:chord_exclusion}.

The star $\gS$ and $\gMP{2,4}$ have
five and eight links, respectively.
Define
$
L_6=\frac23\Ben(\gS)+\frac13\Ben(\gMP{2,4}).
$
For $\gA{6,1}$,
\[
\Ben(\gA{6,1})-L_6
=
\frac23p^3\delta^2\rev{(-q)}
(p^2-3p+9)<0.
\]
For $\gA{6,2}$,
\[
\begin{aligned}
\Ben(\gA{6,2})-L_6
=
\frac23p^3\delta^2
\bigl[
&6\delta p(2-p)\\
&+p^3-4p^2+6p-9
\bigr]<0,
\end{aligned}
\]
since the expression in brackets is increasing
in $\delta$ and, at $\delta=1$, equals
$\rev{-q}(p^2-9p+9)<0$.
Hence neither network can be welfare-optimal.

For the seven-link networks, define
$L_7=\frac13\Ben(\gS)+\frac23\Ben(\gMP{2,4}).$
For $\gA{7,1}$,
\[
\begin{aligned}
\Ben(\gA{7,1})-L_7
=
\frac23p^3\delta^2
\bigl[
&3\delta p(p^2-3p+3)\\
&+2p^3-5p^2+6p-6
\bigr]<0.
\end{aligned}
\]
Indeed, the expression in brackets is increasing
in $\delta$, and its value at $\delta=1$ is
$\rev{-q}(5p^2-9p+6)<0$.
For $\gA{7,2}$, we obtain
\[
\Ben(\gA{7,2})-L_7
=
\frac23\delta^2p^3\rev{(-q)}Q(p,\delta).
\]
Since \rev{$q>0$}, this difference is strictly
negative whenever $Q(p,\delta)>0$.

Finally, consider $\gA{10}$.
The complete multipartite networks
$\gMP{3,3}$ and $\gMP{3,2,1}$ have nine
and eleven links, respectively.
The corresponding chord gives
\[
\begin{aligned}
&\Ben(\gA{10})
-\frac12\left[
\Ben(\gMP{3,3})
+\Ben(\gMP{3,2,1})
\right]\\
&\qquad=-p^3\delta^2\rev{q^{3}}<0.
\end{aligned}
\]
Hence $\gA{10}$ cannot be welfare-optimal.
Under $Q(p,\delta)>0$, all seven identified
non-complete-multipartite networks are
therefore excluded from welfare optimality.
This completes the proof.
\end{proof}

\begin{proof}[\rev{Proof of Proposition \ref{prop:complexity}}]
There are $\mathrm{p}(n)$ integer partitions of $n$.
Each partition $a=(a_1,\ldots,a_r)$ can be evaluated
in $O(r)$ operations using \eqref{eq:cmpL} and
\eqref{eq:cmpB}. Since $r\leq n$, enumerating and
evaluating all partitions requires
$O\bigl(n\,\mathrm{p}(n)\bigr)$ operations.
For each feasible link count, the algorithm retains
the highest benefit among the partitions evaluated.
Since there are at most $\binom{n}{2}+1=O(n^2)$
possible link counts, the resulting frontier can be
ordered and its upper concave envelope constructed
by a monotone sweep in $O(n^2)$ operations.
The overall running time is therefore
$O\bigl(n\,\mathrm{p}(n)+n^2\bigr)$.
Moreover, the algorithm necessarily enumerates all
$\mathrm{p}(n)$ partitions. Hence, by the
\rev{asymptotic formula of Hardy and Ramanujan}, its running
time is $\exp\bigl(\Theta(\sqrt n)\bigr)$.
\end{proof}

\begin{proof}[\rev{Proof of Theorem \ref{thm:dp}}]
Consider a feasible state $(j,s)$. Any multiset
of positive cell sizes satisfying
$\sum_t a_t=j$ and $\sum_t a_t^2=s$
can be decomposed into a cell of size $m$
and a residual multiset satisfying
\[
\sum_t a_t=j-m,
\qquad
\sum_t a_t^2=s-m^2.
\]
Since the objective is additive across cells,
the maximum contribution of the residual multiset
is $\Phi(j-m,s-m^2)$.
Maximizing over all admissible cell sizes $m$
therefore yields \eqref{eq:dprec}.
The first state variable $j$ can take the $n+1$ values
$0,\ldots,n$, while the second state variable $s$ can take
the $n^2+1$ values $0,\ldots,n^2$. Hence, the dynamic program
has at most $(n+1)(n^2+1)=O(n^3)$
states $(j,s)$. For each state, the recursion considers at
most $n$ possible cell sizes. Therefore, evaluating all states
requires $O(n^4)$ arithmetic operations.
For every feasible $s$, the definition of
$\Phi(n,s)$ and the benefit representation
above yield \eqref{eq:dpfrontier}.
Once the restricted frontier points are ordered by link count,
their upper concave envelope can be constructed in
linear time in the number of frontier points, and hence
in $O(n^2)$ operations. Thus, the overall complexity
remains $O(n^4)$.
\end{proof}

\begin{proof}[\rev{Proof of Proposition \ref{prop:underest}}]
\begin{confblock}
For a network $g$ and $u\in(0,1)$, write $\Ben(g;u)$ for the expected benefit of $g$
when each node survives with probability $u$. We first show that $\Ben(g;u)$ is
nondecreasing in $u$, and strictly increasing when $g$ is nonempty.
Let $U_1,\ldots,U_n$ be independent random variables, each uniformly distributed on
$[0,1]$, and set $\hat S=\{i:U_i\le\hat p\}$ and $S=\{i:U_i\le p\}$. Then $\hat S$ and
$S$ are distributed as the sets of surviving nodes under $\hat p$ and $p$, respectively,
and $\hat S\subseteq S$. Since $g|_{\hat S}\subseteq g|_{S}$, any two nodes of $\hat S$
are weakly closer in $g|_{S}$ than in $g|_{\hat S}$. Hence every ordered pair of distinct
nodes generates a weakly larger benefit in state $S$ than in state $\hat S$, where a pair
generates no benefit in a state in which one of its nodes has failed. Taking expectations
pair by pair gives $\Ben(g;\hat p)\le\Ben(g;p)$. Moreover, two linked nodes are at
distance one whenever both survive, so the two ordered pairs of each link contribute
exactly $2u^{2}\delta$ to $\Ben(g;u)$. This contribution is strictly increasing in $u$,
so that $\Ben(g;\hat p)<\Ben(g;p)$ when $g$ is nonempty and $\hat p<p$.

Since link costs do not depend on node survival, it follows that
$\W(g;\hat p,\delta,c)\le\W(g;p,\delta,c)$ for every network $g$, with strict inequality
when $g$ is nonempty and $\hat p<p$. Taking $g=\bar g^{\mathrm{mp}}(\hat p)$ gives the
first inequality in \eqref{eq:underest}. The second holds because the empty network is
complete multipartite and has zero welfare, so that the optimal welfare under $\hat p$
is nonnegative.

Suppose now that the optimal welfare at $p$ is positive. Then $L(\hat p,p)\ge0$, because
$\bar g^{\mathrm{mp}}(p)$ maximizes welfare at $p$ among complete multipartite networks,
and $L(\hat p,p)\le100$ by \eqref{eq:underest}. Finally, let $\hat p<p$. If
$\bar g^{\mathrm{mp}}(\hat p)$ is the empty network, its welfare at $p$ is zero and
$L(\hat p,p)=100$. Otherwise, the strict inequality above gives
$\W(\bar g^{\mathrm{mp}}(\hat p);p,\delta,c)>0$, and hence $L(\hat p,p)<100$.
\end{confblock}
\end{proof}

\subsection*{\rev{Proofs for Section \ref{sec:stability}}}

\begin{proof}[\rev{Proof of Proposition \ref{prop:noadd}}]
Adding $ij$ moves $j$ from distance two to distance one for $i$,
changing its contribution from
$p^{2}\delta^{2}(1-q^{n-a_t})$ to $p^{2}\delta$,
which gives \eqref{eq:addgain}. No other distance changes\rev{, because}
agents outside cell $t$ were already at distance one, and another
member $\rev{z}$ of cell $t$ remains at distance two from $i$, since
$j$ is not adjacent to $\rev{z}$. The expression in \eqref{eq:addgain}
is increasing in $a_t$ because $q^{n-a_t}$ increases with
$a_t$ for $0<q<1$. Hence the largest non-singleton cell
determines the binding condition.
\end{proof}

\begin{proof}[\rev{Proof of Proposition \ref{prop:nosever}}]
Three groups of pairs are affected. First, $j$ itself. After the
deletion the common neighbors of $i$ and $j$ are exactly the $w$
agents outside both cells, so they are at distance two when at
least one of these survives, an event of probability $1-q^{\,w}$.
If none survives, $i$ can still reach $j$ by the three-step route
through a surviving member of $A_u\setminus\{j\}$ and then a
surviving member of $A_t\setminus\{i\}$, which requires both to
be nonempty of survivors, of probability
$(1-q^{\,\rev{y}-1})(1-q^{\,\rev{x}-1})$\rev{. Otherwise} they are disconnected.
This yields the first bracket.

Second, the $\rev{x}-1$ other members of $i$'s own cell. Before the
deletion $i$ reaches each at distance two whenever any of the
$n-\rev{x}$ agents outside cell $t$ survives\rev{. Afterward} the relevant
set has $n-\rev{x}-1$ members, and if none of them survives $i$ is
isolated. The loss per such agent is
$p^{2}\delta^{2}(q^{\,n-\rev{x}-1}-q^{\,n-\rev{x}})
=p^{3}\delta^{2}q^{\,n-\rev{x}-1}$,
giving the second bracket.

Third, every other agent outside cell $t$ remains adjacent to
$i$, so those terms are unchanged. The lower bound follows
since $i$ forfeits $p^{2}\delta$ and recovers at most
$p^{2}\delta^{2}$, the remaining terms being nonnegative.
\end{proof}

\begin{proof}[\rev{Proof of Proposition \ref{prop:separation}}]
Since $c_{\mathrm{mid}}<c_{\mathrm{high}}$ and
$c_{\mathrm{high}}
=p^2\delta+\frac{n-2}{2}p^3\delta^2
>p^2\delta$,
the stated interval is nonempty.
By \rev{Theorem} \ref{thm:star}, throughout this
interval the unique welfare-optimal network
is the star. Hence,
$\Eff(p,\delta,c)=\{\gS\}$,
so that $|\Eff(p,\delta,c)|=1$.
In particular, $\Eff(p,\delta,c)$ is nonempty.

Consider the empty network $\gE$, for which
$\W(\gE)=0$. Adding any link $ij$ yields
$\W(\gE+ij)-\W(\gE)=2(p^2\delta-c)<0$,
since $c>p^2\delta$.
As the link affects only its two endpoints,
which receive identical benefits and incur
identical costs, we have
$\Pi_i(\gE+ij)-\Pi_i(\gE)
=\Pi_j(\gE+ij)-\Pi_j(\gE)
=p^2\delta-c<0$.
Hence, neither endpoint wishes to form the link.
Since there are no existing links to sever,
$\gE$ is pairwise stable, implying that
$\PS(p,\delta,c)$ is nonempty.

It remains to show that the two sets are
disjoint. Consider any welfare-optimal star
$\gS$, and let $h$ denote its center.
If $h$ severs a link $hi$ with a peripheral
agent $i$, its expected benefit decreases
by $p^2\delta$, while it saves the cost $c$.
Therefore,
\[
\Pi_h(\rev{\gS}-hi)-\Pi_h(\rev{\gS})
=c-p^2\delta>0.
\]
Hence, no star is pairwise stable.

Since the star is the unique welfare-optimal
network, we conclude that
$\Eff(p,\delta,c)\cap\PS(p,\delta,c)
=\rev{\emptyset}$.
\end{proof}

\begin{proof}[\rev{Proof of Theorem \ref{thm:existence}}]
Fix $\rev{\eta}\in\{1,\ldots,n-1\}$ and let $(\rev{x},\rev{y})$ be the sizes
of two distinct cells of $g_{[\rev{\eta}]}$, so that
$\max(\rev{x},\rev{y})=\rev{\eta}$ and hence
\begin{equation}\label{eq:minmax}
\rev{x}+\rev{y}-\rev{\eta}-1=\min(\rev{x},\rev{y})-1\ge0.
\end{equation}
Write $w=n-\rev{x}-\rev{y}$. By Proposition \ref{prop:noadd},
$\alpha(g_{[\rev{\eta}+1]})
=p^{2}[\delta-\delta^{2}+\delta^{2}q^{\,n-\rev{\eta}-1}]$,
and by \eqref{eq:severloss},
\[
\Lambda(\rev{x},\rev{y})
=p^{2}\delta-p^{2}\delta^{2}(1-q^{\,w})
-p^{2}\delta^{3}q^{\,w}
(1-q^{\,\rev{x}-1})(1-q^{\,\rev{y}-1})
+(\rev{x}-1)p^{3}\delta^{2}q^{\,n-\rev{x}-1}.
\]

Subtracting $p^2\delta-p^2\delta^2$ from both
sides and dividing by $p^2\delta^2>0$,
the inequality
$\alpha(g_{[\rev{\eta}+1]})\le\Lambda(\rev{x},\rev{y})$
is equivalent to
\[
q^{\,n-\rev{\eta}-1}
\le
q^{\,w}
-\delta q^{\,w}(1-q^{\,\rev{x}-1})(1-q^{\,\rev{y}-1})
+(\rev{x}-1)p\,q^{\,n-\rev{x}-1}.
\]
Now $n-\rev{\eta}-1-w=\rev{x}+\rev{y}-\rev{\eta}-1\ge0$ by
\eqref{eq:minmax} and $n-\rev{x}-1-w=\rev{y}-1$,
so dividing through by $q^{\,w}$ gives
\[
q^{\,\rev{x}+\rev{y}-\rev{\eta}-1}
\le
1-\delta(1-q^{\,\rev{x}-1})(1-q^{\,\rev{y}-1})
+(\rev{x}-1)p\,q^{\,\rev{y}-1}.
\]
By \eqref{eq:minmax}, the left-hand side is
$q^{\,\mu-1}$ with $\mu=\min(\rev{x},\rev{y})$.
Since the final term is nonnegative, it suffices that
$\delta(1-q^{\,\rev{x}-1})(1-q^{\,\rev{y}-1})
\le1-q^{\,\mu-1}$.
Writing $\nu=\max(\rev{x},\rev{y})$,
\[
\delta(1-q^{\,\rev{x}-1})(1-q^{\,\rev{y}-1})
=\delta(1-q^{\,\mu-1})(1-q^{\,\nu-1})
\le\delta(1-q^{\,\mu-1})
\le1-q^{\,\mu-1},
\]
because $1-q^{\,\nu-1}\le1$ and $\delta<1$.
This proves
$\alpha(g_{[\rev{\eta}+1]})\le\Lambda(\rev{x},\rev{y})$
for every pair, hence \eqref{eq:chain}.

For the covering, $\alpha(g_{[\rev{\eta}]})$ is
\rev{strictly increasing} in $\rev{\eta}$, and
$\alpha(g_{[\rev{\eta}]})\le\alpha(g_{[\rev{\eta}+1]})
\le\sigma(g_{[\rev{\eta}]})$
for $\rev{\eta}=1,\ldots,n-1$.
Thus consecutive stability windows meet or overlap.
Since $\alpha(g_{[1]})=0$, the first window starts
at zero. At the other end, $g_{[n]}\cong\gE$ has
$\alpha(g_{[n]})=p^{2}\delta$ and, by
Remark \ref{rem:empty},
$\sigma(g_{[n]})=+\infty$,
since there is no pair of distinct cells and hence
no deletion constraint. Therefore the final window
extends to infinity, and
\[
\bigcup_{\rev{\eta}=1}^{n}
[\alpha(g_{[\rev{\eta}]}),\sigma(g_{[\rev{\eta}]})]=[0,\infty).
\]
Since pairwise stability is invariant under relabeling,
the existence of a stable quotient network implies
that $\mathcal Q^{\mathrm{mp}}(\N)$ contains a
\rev{pairwise-stable} network. This proves the theorem.
\end{proof}


\end{document}